\documentclass[a4paper,11pt]{article}
\pdfoutput=1 % if your are submitting a pdflatex (i.e. if you have
\usepackage{jheppub} % for details on the use of the package, please
\usepackage[T1]{fontenc}
\usepackage[italian,english]{babel}
\usepackage{hyperref}
\usepackage{ifpdf}
\usepackage{subfigure}
\usepackage{amssymb}
\usepackage{amsfonts}
\usepackage{epsf}
\usepackage{rotating}
\usepackage{graphicx}
\usepackage{amsmath}
\usepackage{fancyhdr}
\usepackage{lineno}
\usepackage{babel}
\usepackage{graphics}
\usepackage{pstricks}
\usepackage{color}
\usepackage{multirow}
\usepackage{hhline}
\newcommand{\nn}{\nonumber}
\newcommand{\lsim}{\mathrel{\mathop{\kern 0pt \rlap
  {\raise.2ex\hbox{$<$}}}
  \lower.9ex\hbox{\kern-.190em $\sim$}}}
\newcommand{\gsim}{\mathrel{\mathop{\kern 0pt \rlap
  {\raise.2ex\hbox{$>$}}}
  \lower.9ex\hbox{\kern-.190em $\sim$}}}

\newcommand{\be}{\begin{equation}}
\newcommand{\ee}{\end{equation}}
\newcommand{\bea}{\begin{eqnarray}}
\newcommand{\eea}{\end{eqnarray}}

\def\nto{\not\!\!{\to}}
\title{\boldmath Distinguishing charged Higgs bosons from different representations at the LHC}
\preprint{ IITH-PH-0003/17}
\author[a]{Priyotosh Bandyopadhyay}

\author[b]{Antonio Costantini}

\affiliation[a]{Indian Institute of Technology Hyderabad, Kandi,  Sangareddy-502287, Telengana, India}

\affiliation[a]{Dipartimento di Matematica e Fisica "Ennio De Giorgi", \\ Universit\`a del Salento and INFN-Lecce, \\ Via Arnesano, 73100 Lecce, Italy}

\emailAdd{bpriyo@iith.ac.in}
\emailAdd{antonio.costantini@le.infn.it}

\abstract{Extending the Standard Model (SM) scalar sector via one or multiple Higgs field(s) in higher representation brings one or more charged Higgs bosons in the spectrum.
Some of these gauge representations with appropriate hypercharge can bring up doubly charged Higgs boson and can be easily distinguished from the existing models with only singly charged Higgs boson. In this study we focus on distinguishing the singly charged Higgs bosons from different representations, viz. doublets and triplets of $SU(2)_L$ gauge group. We consider  a  supersymmetric extension of SM with a gauge singlet and $SU(2)_L$ triplet with $Y=0$  as a benchmark scenario with the possibility of rich phenomenology due to existence of light pseudoscalar for $Z_3$ symmetric superpotential. A detailed collider simulation considering all the SM backgrounds has been carried out in order to classify the final states which are favourable to charged Higgs boson from one particular representation than others. 
 We show that such different representations can be probed an distinguished via looking at single charged Higgs boson phenomenology at the LHC with 14 TeV center of mass energy within $\sim 50$ fb$^{-1}$ of integrated luminosity.}

\begin{document}

\maketitle
\flushbottom

\section{Introduction}

In 2012 the ATLAS \cite{ATLASdisc} and CMS \cite{CMSdisc} collaborations announced the discovery of a new elementary particle which was the candidate Higgs boson. This particle was the missing part of the Standard Model (SM) of particle physics. The ATLAS and CMS collaborations found a 125 GeV mass resonance with the properties that are mostly of the SM Higgs boson. The role of the Higgs boson in the SM is to give mass to all the elementary particles, apart from the photon and the gluon, through  the mechanism of spontaneous symmetry breaking (SSB). The discovery made at the Large Hadron Collider (LHC) has revealed the SSB mechanism is realised  in a gauge theory such as the SM by at least one Higgs doublet. However, the possible existence of other scalar bosons cannot be excluded. This is mainly due to the not yet measured self interactions of the Higgs boson. Clearly, they are essential in order 
to establish the mechanism of electroweak symmetry breaking (EWSB), which is crucial in the SM dynamics, with better precision. 

In spite of its success in the explanation of the proprieties of the known elementary particles, the SM is not a completely satisfactory theory. There are, in fact, long-standing issues, such that the gauge-hierarchy problem, the cold dark matter candidate, the masses of the neutrinos etc., to which the SM does not provide a satisfactory answer. 

In this article we are interested in extensions of Higgs sectors via various possible gauge 
representations. In principle these extensions are possible both with and without supersymmetry but in this analysis we consider only the supersymmetric ones. However the Higgs phenomenology and
the non-standard decays are similar in both the cases. The interesting fact is that any extension of SM other than with a singlet superfield which takes part in EWSB gives rise to at least one physical massive charged Higgs boson. In the case of minimal supersymmetric extension of SM (MSSM) we have two Higgs doublets with opposite hypercharge which results in a physical charged Higgs bosons $h^\pm$. We refer to such a charged Higgs boson  as a doublet-type charged Higgs boson and its coupling to SM fermions are given by Yukawa couplings and $\tan{\beta}$, the ratio of the vevs of the two Higgs doublets \cite{anatomyII}.

The other possibilities come from the triplet representation of SU(2). It is possible to include triplet(s) with $Y=0$, $Y=\pm 1$ or both \cite{notetrip}. Each of them has its own signature, apart from addition of one or more charged Higgs bosons to the spectrum. The simplest extension is with a $Y=0$ triplet which gives rise to two more physical charged Higgs boson after the EWSB \cite{TESSM, TNMSSM}. Such extension is constrained by the $\rho$ parameter \cite{rho} because the SU(2) triplet with $Y=0$ hypercharge contributes to the tree-level mass of the $W^\pm$ gauge boson but not to the $Z$ one. The vev of the $Y=0$ triplet, which breaks the custodial symmetry, is then restricted to $\lesssim 5$ GeV \cite{TESSM, TNMSSM}. Breaking of custodial symmetry leads to the non-standard signature of $h^\pm \to Z W^\pm$ which is very typical of triplet extension \cite{TESSM, TNMSSM}.

As we already mentioned it is also possible to consider triplet(s) with non-zero hypercharge. Such extension is possible for $Y=\pm 1$, which not only leads to triplet-like charged Higgs bosons but also predicts the existence of doubly charged Higgs boson \cite{agashe}. Even this extension breaks the custodial symmetry and hence the triplet vevs is also restricted by the $\rho$ parameter, similarly to the $Y=0$ scenario. However a combination of $Y=0$ and $Y=\pm 1$ can restore the custodial symmetry which was shown by Georgi-Machacek \cite{GM} and its supersymmetric extension has also been studied \cite{quiros}.

The charged Higgs boson phenomenology can be largely affected by the existence of a very light scalar ($CP$-even or $CP$-odd) viz, next-to-Minimal Supersymmetric Standard Model (NMSSM) \cite{ellwanger}. In this case $h^\pm \to a_1 W^\pm$ is kinematically allowed for a light charged Higgs boson $\lesssim 200$ GeV. In a $Z_3$ symmetric superpotential viz, NMSSM such light pseudoscalar is a pseudo Nambu-Goldstone (pNG) boson of a global $U(1)$ symmetry.  Such $Z_3$ symmetric superpotential can also lead to the existence of similar light pseudoscalar in the triplet extensions also \cite{agashe, TNMSSM}.  Existence of light boson(s) makes the triplet charged Higgs boson phenomenology further more interesting. In this article our main focus is to distinguish triplet-like extensions from the usual doublet-like extensions. We will consider an extension of MSSM with a SM gauge singlet and a $Y=0$ triplet superfields, called the TNMSSM \cite{TNSSMo}. It gives the possibility to probe both the non-standard modes of the light charged Higgs boson decays, \textit{i.e.} $h^\pm\to a_1 W^\pm$ and $h^\pm \to Z W^\pm$. The first one is possible due to existence of pNG boson and the second one is due to triplet structure of the light charged Higgs boson.

The triplet type charged Higgs bosons are different from the doublet-type in two aspects. Firstly they do not couple directly to fermions with the usual Yukawa interaction and secondly they can decay into  $Z\,W^\pm$. So far the searches of the light charged Higgs boson at colliders, specially at the LHC, are done focusing on the doublet-like light charged Higgs only \cite{chargedH}. The doublet-like charged Higgs boson is obtained through fermionic production modes viz., $pp\to t bh^\pm/$, $gb\to t, h^\pm$ and then it is searched via $h^\pm \to \tau \nu/t b$. A triplet-like charged Higgs boson will be surely missed in that case because of the suppressed production and as well as in the decay modes in fermions. In a PYTHIA based analysis we tried to distinguish such light doublet- and triplet-type charged Higgs bosons.

 We organize the paper as follows. In section~\ref{dmodels} we give a brief summer of possible extensions of the Higgs sectors where the 
charged Higgs can come from different representations. Section~\ref{model} gives a summer of supersymmetric extension of SM as TNMSSM. Possible non-standard decays of triplet type charged Higgs bosons are discussed in section~\ref{chb}. In section~\ref{secbps} we discuss the phenomenology of thee triplet and doublet-like light charged Higgs boson and choose some benchmark points for a collider study at typical LHC. In section~\ref{sigsim} we perform a detail collider simulation for the signal and consider all the dominant SM backgrounds for the chosen final states, presenting the relative results. In section~\ref{recon} we present the results for the reconstruction of the charged Higgs mass and in section~\ref{disting} the correlation among various model with singly and doubly charged Higgs bosons before our conclusions, which are contained in section~\ref{concl} .

\section{Charged Higgs bosons in various supersymmetric extensions of the SM}\label{dmodels}
In the Standard Model we do not have any physical charged Higgs boson and it can be achieved  by extending to scalar sectors by adding at least one more $SU(2)_L$ doublet or triplet, that take part in electroweak symmetry breaking.  
We summarize in this section the status of the charged Higgs sector in various theory beyond the SM, comparing the number of the charged Higgs bosons and their most important features in terms of allowed decay modes. In Table \ref{cHdt} we present the total Higgs spectrum along with the most important feature of the lightest charged Higgs boson in various supersymmetric extensions of the SM.

\begin{table}
\begin{center}
\renewcommand{\arraystretch}{1.5}
\begin{tabular}{||c|c|c||}
\hline\hline
MSSM {\bf(a)}&NMSSM {\bf(b)}&NMSSM + $A_i\sim0$ {\bf(c)}\\
\hline
\multicolumn{2}{||c|}{$H^\pm,\,A$ degenerate}&$H^\pm\to a_1W^\pm$ allowed\\
\multicolumn{2}{||c|}{1 $H^\pm$}&1 $H^\pm$\\
\hline\hline
\multicolumn{3}{||c||}{Triplet Superfield $\hat T$}\\
\hline
$Y=0$ {\bf (d)}&$Y=\pm1$ {\bf(e)}&$Y=0,\pm1$ {\bf(f)}\\
\hline
$h_1^\pm\nto\, a_1W^\pm$  ($m_{a_1}\gsim m_{h_1^\pm}$)&$h_1^\pm\to ZW^\pm$ allowed&custodial symmetry\\
3 $h_i^\pm$&3 $h_i^\pm$, 2 $h_j^{\pm\pm}$&5 $h_i^\pm$, 2 $h_j^{\pm\pm}$\\
\hline\hline
\multicolumn{3}{||c||}{MSSM + $\hat S$ + $\hat T\;\rm{with}\;Y=0$ + $A_i\sim0$ {\bf(g)}}\\
\hline
\multicolumn{3}{||c||}{$h_1^\pm$ triplet-type}\\
\multicolumn{3}{||c||}{$h_1^\pm\to Z\,W^\pm$ enhanced for $\lambda_T\sim0$}\\
\multicolumn{3}{||c||}{$h_1^\pm \to a_1W^\pm$ allowed}\\
\multicolumn{3}{||c||}{3 $h_i^\pm$}\\
\hline\hline
\end{tabular}
\caption{The charged Higgs sector in various SUSY theories.}\label{cHdt}
\end{center}
\end{table}

The MSSM is the most simple supersymmetric theory which allows a single charged Higgs boson in the spectrum. However the charged Higgs boson decays mostly in fermionic modes viz, $\tau, \nu$ and/or $t, b$. There is an inherent mass degeneracy (nearly) with the pseudoscalar boson ($A$) and heavy Higgs boson ($H$) present in the spectrum which prohibits the decays of $h^\pm \to A/H W^\pm$ as shown in scenario {\bf(a)}. In the simplest extension of MSSM, \textit{i.e.}  NMSSM the situation is similar, scenario {\bf(b)}. This is due to the fact that the singlet superfield, after acquiring vev, give rise to an additional scalar and pseudo-scalar in the spectrum but not a charged Higgs boson. For an inert singlet (which does to take part in EWSB) can give rise to a charged Higgs boson which is SM gauge singlet. The degeneracy between the charged Higgs boson and the lightest pseudoscalar (scenario {\bf(b)}, {\bf(a)}) can be easily lifted by considering $Z_3$ symmetric superpotential (scenario {\bf(c)}) which is realized in the limit $A_i\to0$. In this limit there is an extra U(1) global symmetry, known as the R-symmetry in the literature \cite{ellwanger}, which is common to supersymmetric theories with cubic terms in the superpotential. The lightest pseudoscalar of the spectrum is the Nambu-Goldstone mode of this extra symmetry, if $A_i\equiv0$. If the symmetry is softly broken the pseudoscalar takes the role of an axion-like particle. Having a very low mass  the decay $H^\pm \to a_1 W^\pm$ is kinematically allowed, however such light pseudoscalar faces direct and indirect constraints from LEP \cite{LEPb} and other experiments \cite{bpys}.

Scenarios {\bf (d)}, {\bf (e)} and {\bf (f)} correspond to $Y=0$, $Y=\pm1$ and  $Y=0,\pm1$ (Georgi-Machacek) cases where one or more $SU(2)_L$ triplet superfields are added to the MSSM superfield content. In the simplified extension of  {\bf (d)}, where a $Y=0$  triplet superfield is added to the MSSM, there are two more singly charged Higgs boson in the spectrum respect to the MSSM. The triplet charged Higgs bosons do not couple directly to the fermions, which makes their production at the LHC rather difficult  and they decay in non-standard modes \cite{pbas3}.  Due to this reason for the triplet-like charged Higgs boson, the decay to fermions is suppressed. In this scenario the two-body decay $h_1^\pm\to a_1W^\pm$ is not allowed because $m_{h_1^\pm}\lesssim m_{a_1}$ \cite{pbas3}. The most important decay mode is then $h_1^\pm\to ZW^\pm$. This interaction is present at tree-level in theories with scalar triplets which acquire vevs and break custodial symmetry at the tree-level. Instead, if the model has scalar doublets or singlets of $SU(2)_L$ the breaking of the custodial symmetry is only possible at loop level and so is the decay $h_i^\pm\to Z\,W^\pm$. If the triplet superfield has $Y=\pm1$ hypercharge, (scenario {\bf (e)}), then in the spectrum, we will see a doubly charged Higgs bosons along with two additional triplet type singly charged Higgs bosons \cite{agashe}. Finally scenario {\bf (f)} has both a $Y=0$ and a $Y=\pm1$ triplet superfields and the corresponding charged Higgs bosons. This is the supersymmetric version of the well-known Georgi-Machacek model \cite{GM}. The most important feature of scenario {\bf(f)} is the custodial symmetry which can be naturally imposed \cite{quiros}.

Scenario {\bf (g)} is considered in \cite{OurCharged}, where on the top of the MSSM superfield content there is a singlet superfield and a triplet superfield with $Y=0$. The superpotential is $Z_3$ symmetric, that means that only cubic terms are allowed. In the limit $A_i\sim0$ it exhibit a softly broken U(1) symmetry, similarly to scenario {\bf (c)} and the lightest pseudoscalar is the pseudo Nambu-Goldstone mode of this extra $U(1)$, very low in mass. In this scenario the lightest charged Higgs boson can decay in $a_1 W^\pm$ and $Z\,W^\pm$ and if it is triplet-like its decay into fermions is suppressed. In the section below we briefly introduce the model which will be used later for the phenomenological studies to distinguish among doublet and triplet charged Higgs boson(s) at the LHC.

\section{The Model}\label{model}

We consider an extension of the MSSM with a (gauge) singlet superfield $\hat S$ and a triplet superfield $\hat T$ with $Y=0$. The model is detailed in \cite{TNSSMo} and here we will give the very basic proprieties. The Higgs superfields are given below,

\begin{equation}\label{spf}
 \hat T = \begin{pmatrix}
       \sqrt{\frac{1}{2}}\hat T^0 & \hat T_2^+ \cr
      \hat T_1^- & -\sqrt{\frac{1}{2}}\hat T^0
       \end{pmatrix},\qquad \hat{H}_u= \begin{pmatrix}
      \hat H_u^+  \cr
       \hat H^0_u
       \end{pmatrix},\qquad \hat{H}_d= \begin{pmatrix}
      \hat H_d^0  \cr
       \hat H^-_d
       \end{pmatrix},\qquad \hat S,
 \end{equation}

where $\hat T^0$ is a complex neutral superfield, while  $\hat T_1^-$ and $\hat T_2^+$ are the charged Higgs superfields. $\hat{H}_u$ and $\hat{H}_d$ are the usual doublet superfields of the MSSM and $\hat S$ is the singlet superfiled. The gauge symmetry implies that the Yukawa interactions are identical to the MSSM ones, because neither the singlet nor the triplet superfields have any interaction with the fermionic superfields. This means that the superpotential can be written as 

 \begin{equation}
 W_{TNMSSM}=W_{MSSM} + W_{TS},
 \end{equation}
with
\begin{equation}
W_{MSSM}= y_t \hat U \hat H_u\!\cdot\! \hat Q - y_b \hat D \hat H_d\!\cdot\! \hat Q - y_\tau \hat E \hat H_d\!\cdot\! \hat L\ ,
\label{spm}
 \end{equation}
 being the superpotential of the MSSM, while 
 \begin{equation}
W_{TS}=\lambda_T  \hat H_d \cdot \hat T  \hat H_u\, + \, \lambda_S \hat S  \hat H_d \cdot  \hat H_u\,+ \frac{\kappa}{3}\hat S^3\,+\,\lambda_{TS} \hat S  \textrm{Tr}[\hat T^2]
\label{spt}
 \end{equation}
Here ''$\cdot$'' denotes a contraction with the Levi-Civita symbol $\epsilon^{ij}$, with $\epsilon^{12}=+1$.

It is a characteristic of any scale invariant supersymmetric theory with a cubic superpotential that the complete Lagrangian with the soft SUSY breaking terms has an accidental  $Z_3$ symmetry. This is generated by the invariance of all of its components after multiplication of the chiral superfields by the phase $e^{2\pi i/3}$. This $Z_3$ symmetry is promoted to a global $U(1)$ symmetry in the limit $A_i\to0$, where the $A_i$ are the trilinear terms of the soft-breaking part of the scalar potential \cite{TNSSMo}. This global $U(1)$ symmetry can be softly broken by small $A_i$, giving rise to a very light pseudoscalar which is the pseudo Nambu-Goldstone mode of the symmetry.

 We assume that all the coefficients involved in the Higgs sector are real in order to preserve CP invariance. The breaking of the $SU(2)_L\times U(1)_Y$ electroweak symmetry is then obtained by giving real vevs to the neutral components of the Higgs field
 \be
 <H^0_u>=\frac{v_u}{\sqrt{2}}, \, \quad \, <H^0_d>=\frac{v_d}{\sqrt{2}}, \quad ,<S>=\frac{v_S}{\sqrt{2}} \, \quad\, <T^0>=\frac{v_T}{\sqrt{2}},
 \ee
 which give mass to the $W^\pm$ and $Z$ bosons
 \be
 m^2_W=\frac{1}{4}g^2_L(v^2 + 4v^2_T), \, \quad\ m^2_Z=\frac{1}{4}(g^2_L \, +\, g^2_Y)v^2, \, \quad v^2=(v^2_u\, +\, v^2_d), 
\quad\tan\beta=\frac{v_u}{v_d} \ee
 and also induce, as mentioned above, a $\mu$-term of the form $ \mu_D=\frac{\lambda_S}{\sqrt 2} v_S+ \frac{\lambda_T}{2} v_T$.
 The triplet vev $v_T$ is strongly  constrained by the global fit on the measurement of the $\rho$ parameter \cite{rho}
 \be
 \rho =1.0004^{+0.0003}_{-0.0004} ,
 \ee 
 which restricts its value to $v_T \leq 5$ GeV. Respect to the tree-level expression, the non-zero triplet contribution to the $W^\pm$ mass leads to a deviation of the $\rho$ parameter
 \be
 \rho= 1+ 4\frac{v^2_T}{v^2} .
 \ee

\section{Triplet-like singly Charged Higgs bosons}\label{chb}

The lightest triplet-like charged Higgs in TNMSSM \cite{TNSSMo, OurCharged, TNSSMlps} can decay to $Z\,W^\pm$ as well as to $a_1W^\pm$. Establishing these two non-standard decay modes will be sufficient to prove the existence of higher representations $SU(2)_L$ in the Higgs spectrum, \textit{i.e.} the triplet as well as the existence of  SM gauge singlet in the spectrum. As we already point out in section \ref{dmodels}, in the case of a triplet with non-zero hypercharge, there is a doubly-charged Higgs boson in the spectrum. Its phenomenology have been studied extensively in the literature \cite{nonzerhy}. In this article our goal is to distinguish between doublet- and triplet-like charged Higgs boson by searching for singly charged Higgs boson at the LHC in appropriate decay channels.

The phenomenology of the lightest charged Higgs boson of the TNMSSM is affected by the presence of a light pseudoscalar, which induces a new decay mode. Along with the existence of the light pseudoscalar, which opens up the $h^\pm_1 \to a_1 W^\pm$ decay channel, the triplet-like charged Higgs boson adds another tree-level decay mode, not possible otherwise. In particular, a $Y=0$ triplet-like charged Higgs boson can decay into $Z\,W^\pm$, which is a signature of custodial symmetry breaking. If the model has only doublets of $SU(2)_L$ then the $h^\pm\to Z\, W^\pm$ decay is a loop-induced one. Apart from that, the usual doublet-like decay modes into $\tau\nu$ and $tb$ are present via the mixing with the doublets. We summarize the different possible decay modes in the following paragraphs.

The trilinear couplings with charged Higgs bosons, scalar (pseudoscalar) Higgs bosons and $W^\pm$ are given by 
\begin{align}\label{hachW}
g_{h_i^\pm W^\mp h_j}&=\frac{i}{2}g_L\Big(\mathcal R_{j2}^S\mathcal R_{i3}^C-\mathcal R_{j1}^S\mathcal R_{i1}^C+\sqrt2\mathcal R_{j4}^S\left(\mathcal R_{i2}^C+\mathcal R_{i4}^C\right)\Big), \\
g_{h_i^\pm W^\mp a_j}&=\frac{g_L}{2}\Big(\mathcal R_{j1}^P\mathcal R_{i1}^C+\mathcal R_{j2}^P\mathcal R_{i3}^C+\sqrt2\mathcal R_{j4}^P\left(\mathcal R_{i2}^C-\mathcal R_{i4}^C\right)\Big),
\end{align}
where $\mathcal{R}^S, \mathcal{R}^P, \mathcal{R}^C$ are the mixing matrix corresponding to scalar, pseudoscalar and charged Higgs bosons respectively, with $\mathcal R_{j4}^S, \mathcal R_{j4}^P,  \mathcal R_{j,2}^C,  \mathcal R_{j4}^C$ being the triplet part of the mixing matrices \cite{OurCharged}.

Both the triplet and doublet have $SU(2)_L$ charge and hence they couple to $W^\pm$ boson. However the $W^\pm$ boson is in the triplet representation of $SU(2)_L$ and this means that in the $W^\pm\, h_i h_j^\mp$ coupling the neutral and charged Higgs bosons have to be doublet(triplet) and triplet(doublet) type respectively in order to maintain the gauge invariance. 

The decay width of a massive charged Higgs boson in a $W$ boson and a scalar (or pseudoscalar) boson is given by
\bea \label{chwah}
\Gamma_{h_i^\pm\rightarrow W^\pm h_j/a_j}&=&\frac{G_F}{8\sqrt2\pi}m^2_{W^\pm}|g_{h_i^\pm W^\mp h_j/a_j}|^2 \,\sqrt{\lambda(1,x_W,x_{h_j/a_j})}\,\lambda(1,y_{h_i^\pm},y_{h_j/a_j})
\eea
where $x_{W,h_j}=\frac{m^2_{W,h_j}}{m^2_{h_i^\pm}}$ and $y_{h_i^\pm,h_j}=\frac{m^2_{h_i^\pm,h_j}}{m^2_{W^\pm}}$ and similarly for $a_j$. In TNMSSM this decay channel is prominent for a light charged Higgs boson and it is the dominant decay mode if the charged Higgs boson is triplet-like, because of the suppression of the fermionic couplings.

In theories with $Y=0, \pm 2$ hypercharge triplets, which generally break the custodial symmetry, there is a tree-level interactions $h_i^\pm-W^\mp-Z$ \cite{notetrip}. In the TNMSSM this coupling is given by 
\bea\label{zwch}
g_{h_i^\pm W^\mp Z}&=&-\frac{i}{2}\left(g_L\, g_Y\left(v_u\sin\beta\,\mathcal R^C_{i1}-v_d\cos\beta\,\mathcal R^C_{i3}\right)+\sqrt2\,g_L^2v_T\left(\mathcal R^C_{i2}+\mathcal R^C_{i4}\right)\right),
\eea
where the explicit definition of the rotation angles can be found in \cite{OurCharged}. The on-shell decay width is given by
\bea\label{chzw}
\Gamma_{h_i^\pm\rightarrow W^\pm Z}&=&\frac{G_F\,\cos^2\theta_W}{8\sqrt2\pi}m^3_{h_i^\pm}|g_{h_i^\pm W^\mp Z}|^2\,\sqrt{\lambda(1,x_W,x_Z)}\left(8\,x_W\,x_Z+(1-x_W-x_Z)^2\right)\nn\\
\eea
where $\lambda(x,y,z)=(x-y-z)^2-4\,y\,z$ and $x_{Z,W}=\frac{m^2_{Z,W}}{m^2_{h_i^\pm}}$ \cite{Asakawa}. As it is extensively explained in \cite{OurCharged}, this decay channel is enhanced for a triplet-like charged Higgs boson in the limit $\lambda_T\sim0$ due to the same sign values of $\mathcal{R}^C_{12}$ and $\mathcal{R}^C_{14}$.

Beside the non-zero $h^\pm_i-W^\mp-Z$ coupling at the tree-level due to custodial symmetry breaking, the charged Higgs bosons can also decay into fermions through the Yukawa interaction given below
\bea
g_{h_i^+ \bar u d}=i\left(y_u\,\mathcal R^C_{i1}\,\mathtt{P_L}+y_d\,\mathcal R^C_{i3}\,\mathtt{P_R}\right)
\eea
governed by doublet part of the charged Higgses. The decay width at leading order is
\begin{align}\label{chtb}
\Gamma_{h_i^\pm\rightarrow u\,d}&=\frac{3}{4}\frac{G_F}{\sqrt2\pi}m_{h_i^\pm}\sqrt{\lambda(1,x_u,x_d)}\Bigg[(1-x_u-x_d)\,\left(\frac{m^2_u}{\sin^2\beta}(\mathcal R^C_{i1})^2+\frac{m_d^2}{\cos^2\beta}(\mathcal R^C_{i3})^2\right)\nn\\
&\hspace{4.5cm}-4\frac{m_u^2m_d^2}{m^2_{h_i^\pm}}\frac{\mathcal R^C_{i1}\mathcal R^C_{i3}}{\sin\beta\cos\beta}\Bigg]
\end{align}
where $x_{u,d}=\frac{m^2_{u,d}}{m^2_{h_i^\pm}}$. 
The decay of the charged Higgs bosons into quarks is suppressed in the case of triplet-like eigenstates, where $\mathcal{R}^C_{i1,i3}\ll1$.

\section{Benchmark points for the collider study}\label{secbps}
In this section we choose points to perform the collider study and to distinguish doublet- and triplet-like points at the LHC. In Table~\ref{bps} we show the mass spectrum of the selected benchmark points. Together with the recent Higgs data we have also considered the recent bounds on the stop and sbottom masses \cite{thridgensusy} and the mass bounds on the lightest chargino from LEP \cite{chargino}. We have also taken into account the recent bounds on the charged Higgs boson mass from both CMS \cite{ChCMS} and ATLAS \cite{ChATLAS}. These bounds have been derived in their searches for the light charged Higgs bosons from the decay of a top quark, and in decays to $\tau \bar{\nu}$. Addition of any new decay mode will further lower the lower bound for the charged mass exclusion. 

 %%%%%%%%%%%%%%%%% Benchmark points%%%%%%%%%%%%%%%%%%%%
\begin{table}
\begin{center}
\renewcommand{\arraystretch}{1.5}
\begin{tabular}{||c||c|c|c|c||}
\hline\hline
Benchmark&BP1&BP2&BP3 & BP4\\
Points & &&&\\ \hline\hline$m_{h_1}$ & {\color{red}$\sim 125$} & {\color{red}$\sim 125$} & {\color{red}$\sim125$}& {\color{red}$\sim125$}\\
\hline
$m_{h_2}$ &316.14  & \color{green}340.44 &272.87& \color{green}174.21 \\
\hline
$m_{h_3}$& 522.41 & 382.56 & 358.12 &1027.3\\
\hline
$m_{h_4}$ &673.45  & 514.16 & 2094.4 & 1547.7\\
\hline
\hline
$m_{a_1}$ & \color{blue}41.221 & \color{blue}36.145 &  \color{blue}30.655& \color{blue}61.537\\
\hline
$m_{a_2}$ & \color{green}181.34 & 428.68 & 278.22 &1052.7\\
\hline
$m_{a_3}$& 559.32 & 519.01  & 2149.4 & 1325.9\\
\hline
\hline
$m_{h^\pm_1}$ & \color{green}179.69 & \color{green}339.97 &\color{red}289.51& \color{green}{174.11} \\
\hline
$m_{h^\pm_2}$ &316.20  & 399.84  & 2089.7 & 1032.9\\
\hline
$m_{h^\pm_3}$&535.21  & 519.02 & 2144.9  & 1325.9\\
\hline
\hline
\end{tabular}
\caption{Benchmark points for a collider study consistent with the $\sim 125$ GeV Higgs mass, where the $h_{i=1,2,3,4}$, $a_{i=1,2,3}$ are at one-loop and $h^{\pm}_{i=1,2,3}$ masses are calculated at tree level. We color in red the states which are mostly doublets ($>90\%$) and in blue and green those which are mostly singlet and triplet ($>90\%$) respectively. The points are consistent with the $2\sigma$ limits of $h_{125}\to WW^*, ZZ^*, \gamma\gamma$ \cite{CMS, ATLAS}.}\label{bps}
\end{center}
\end{table}
%%%%%%%%%%%%  

%%%%%%%%%%%%%%%%% h_125 decay branching fraction (with tree level mass)%%%%%%%%%%%%%%%%%%%%
\begin{table}
\begin{center}
\renewcommand{\arraystretch}{1.5}
\begin{tabular}{||c||c|c|c|c||}
\hline\hline
Benchmark&\multicolumn{4}{|c||}{Branching ratios}\\
\cline{2-5}
Points& $a_1 a_1$& $W^\pm W^\mp$ & $Z\,Z$  & \;$b\,\bar b$ \;\\
\hline\hline
BP1 &0.105 &0.148&0.020&0.686\\
\hline
BP2 &0.045 &0.143&0.019&0.748\\
\hline
BP3 &0.052 &0.115&0.015&0.770\\
\hline
BP4 & 0.057&0.129&0.017&0.752\\
\hline\hline
\end{tabular}
\caption{Relevant decay branching ratios of $h_{125}$ for the benchmark points.}\label{hdcy2}
\end{center}
\end{table}
%%%%%%%%%%%%

%%%%%%%%%%%%%%%%% a_1 decay branching fraction (with tree level mass)%%%%%%%%%%%%%%%%%%%%
\begin{table}
\begin{center}
\renewcommand{\arraystretch}{1.5}
\begin{tabular}{||c||c|c|c||}
\hline\hline
Benchmark&\multicolumn{3}{|c||}{Branching ratios(\%)}\\
\cline{2-4}
Points& \;$b\bar{b}$ \;&\;$\tau \bar{\tau}$&$\mu\bar\mu$  \\
\hline\hline
BP1 & 0.942   &$5.77\times10^{-2}$ &$2.06\times10^{-4}$ \\
\hline
BP2 & 0.942 & $5.80\times10^{-2}$ &$2.07\times10^{-4}$ \\
\hline
BP3 & 0.941 & $5.85\times10^{-2}$&$2.09\times10^{-4}$ \\
\hline
BP4 & 0.943 & $5.72\times10^{-2}$&$2.03\times10^{-4}$ \\
\hline\hline
\end{tabular}

\caption{Decay branching ratios of $a_1$ for the benchmark points $BP_i$. The kinematically forbidden decays are marked with dashes.}\label{a1dcy2}
\end{center}
\end{table}
%%%%%%%%%%%%

The branching fractions of the SM-like Higgs boson for the four benchmark points are presented in Table \ref{hdcy2}. They are consistent with the observed branching fraction at $2\sigma$ level. The decay channels of the light pseudoscalar $a_1$ are presented in Table \ref{a1dcy2}. The branching fractions for the selected benchmark points are very similar because of the singlet-type selection for the lightest pseudoscalar. Four different benchmark points present different dominant decay modes of the charged Higgs bosons. The mostly triplet-like charged Higgs would dominantly decays into $ZW^\pm$. Similarly for some points it can decay to $a_1 W^\pm$ and for mostly doublet-like charged Higgs boson it decays to $t\, b$.

%%%%%%%%%%%%%%%%% ch_1 decay branching fraction (with tree level mass)%%%%%%%%%%%%%%%%%%%%
\begin{table}
\begin{center}
\renewcommand{\arraystretch}{1.5}
\begin{tabular}{||c||c|c|c|c|c|c||}
\hline\hline
Benchmark&\multicolumn{6}{|c||}{Branching ratios}\\
\cline{2-7}
Points & $a_1 W^\pm$& $h_{125} W^\pm$ & $Z\,W^\pm$  & \;$t\,b$ \;&\;$\tau\, \nu_\tau$&$\mu\,\nu_\mu$ \\
\hline\hline
BP1 &0.969 &$5.56\times10^{-5}$   & $2.09\times10^{-2}$ & $9.58\times10^{-3}$ &$1.29\times10^{-4}$  &$4.57\times10^{-7}$\\
\hline
BP2 &$3.01\times10^{-2}$  & 0.213 & 0.236 & 0.520  & $8.87\times10^{-5}$ &$3.15\times10^{-7}$ \\
\hline
BP3 & $6.43\times10^{-2}$  & $3.51\times10^{-3}$ & $1.94\times10^{-10}$  & 0.932 & $9.25\times10^{-5}$ & $3.28\times10^{-7}$ \\
\hline
BP4 &$1.03\times10^{-5}$ &$3.34\times10^{-3}$ & 0.994  & - &$1.60\times10^{-3}$ &$5.68\times10^{-6}$ \\
\hline\hline
\end{tabular}
\caption{Decay branching ratios of $h_1^\pm$ for the benchmark points. The kinematically forbidden decays are marked with a dash.}\label{chdcy2}
\end{center}
\end{table}
%%%%%%%%%%%%
In Table~\ref{chdcy2} presents the two-body decay branching fractions of the charged Higgs for the benchmark points. We can see that for BP1, the decay branching fraction to $a_1 W^\pm$ is highest.  BP2 seems to have decent decay rates into $h_{125}W^\pm, \, ZW^\pm$ and into $tb$ making it a mixed point where both doublet and triplet natures manifest. In contrary to BP1, BP3 is a doublet-like points which decays mostly to $tb$. Finally, BP4 is characterized by a single dominant decay channel, $h_1^\pm\to Z\,W^\pm$ with a branching fraction of $0.994$. We will see later in the section how we can distinguish such doublet- and triplet-like points. The color code of green is used in Table~\ref{bps} to represent a triplet-like charged Higgs boson and red for the doublet-like. We see that for BP3 the charged Higgs boson is mostly doublet. Similarly blue color has been used to signify singlet-like light pseudoscalar boson. 

\begin{figure}[bth]
	\begin{center}
		\mbox{
				\includegraphics[width=.7\linewidth]{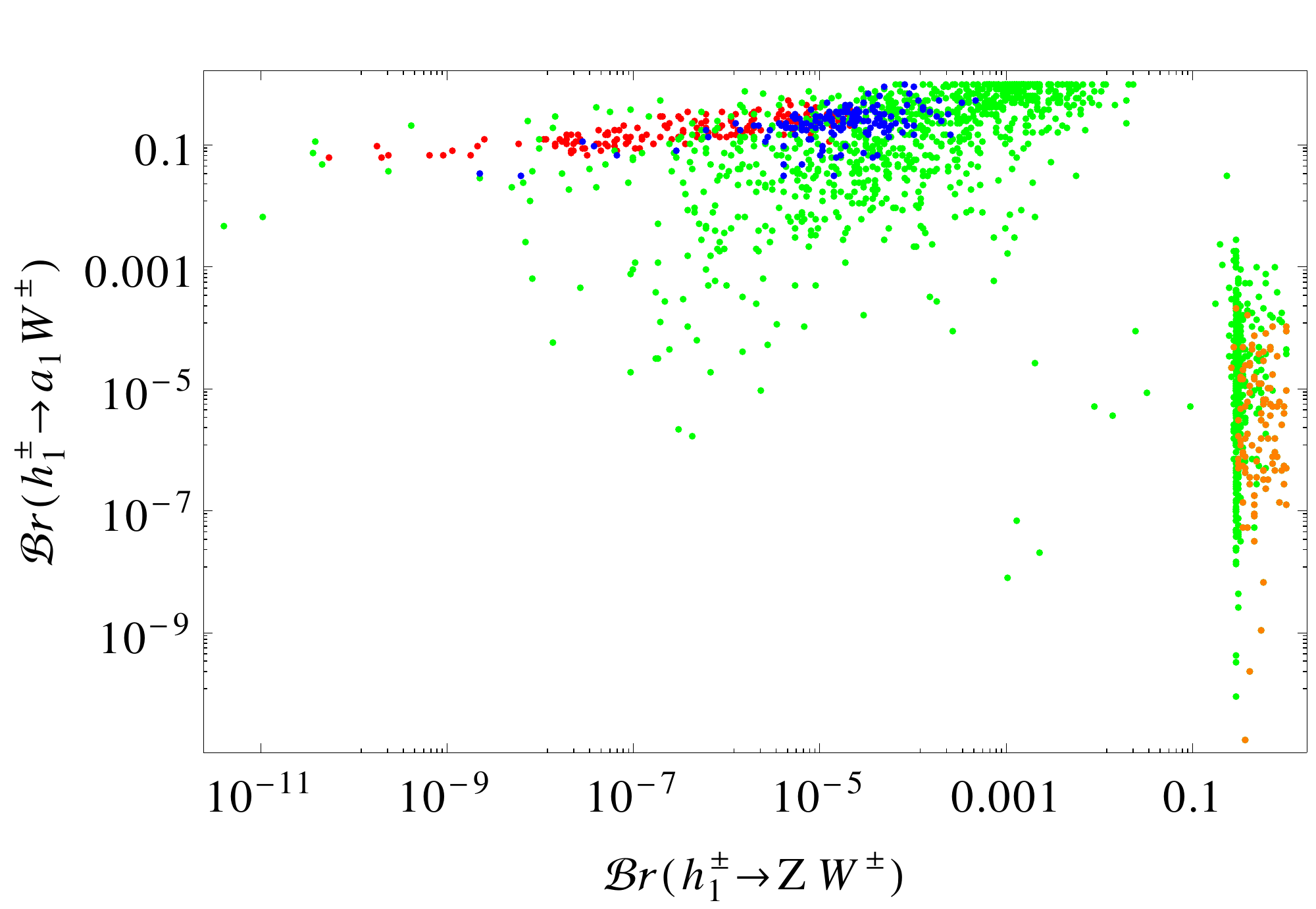}}
		\caption{Correlation plot of the branching ratios for the non-standard decays $h_1^\pm\to Z\,W^\pm$ and $h_1^\pm\to a_1W^\pm$. The red points are doublet-like charged Higgs, the green ones are triplet-like and the blue ones are mixed-one. The orange ones are characterized by $\lambda_T\sim0$ \cite{OurCharged}. We can see that these two decay channels are almost mutually excluded.}\label{brawZW}
	\end{center}
\end{figure}

Figure~\ref{brawZW} describe a correlation plot for $\mathcal{B}r(h^\pm_1 \to a_1 W^\pm)$ and  $\mathcal{B}r(h^\pm_1 \to Z W^\pm)$ for triplet-like light charged Higgs boson ($h^\pm_1$). The red points are doublet-like charged Higgs, the green ones are triplet-like and the blue ones are mixed-one. The orange ones are characterized by $\lambda_T\sim0$ \cite{OurCharged}. We can see that $h^\pm_1 \to a_1 W^\pm$ and 
$h^\pm_1 \to Z W^\pm$ decay channels are almost mutually excluded and it would be very difficult to probe both triplet nature or the charged Higgs boson as well as the existence of the light pseudoscalar $a_1$.
%%%%%%%%%%%%%%%%%%%%%%%%%
%%%%%%%%%%%%%%%%%%%%%%

We calculated the light charged Higgs production cross-sections in pairs and in association of other particles at the LHC for these benchmark points. For this purpose we have implemented the model in SARAH \cite{sarah} and we have generated the model files for CalcHEP \cite{calchep}. The cross-sections have been calculated at the tree-level via Calchep$\_$3.6.23  \cite{calchep} and Table~\ref{Hcrosssec} presents the cross-sections which include the associated K-factors \cite{djouadi}. We can see that for BP1 $h^\pm_1$ and $a_2$ are degenerate and for other BPs, $h^\pm_1$ is degenerate with $h_2$ (see Table~\ref{bps}). The cross-section contributions are also dominant form the respective associated production processes only.

%%%%%%%%%%%%%%%%% h^\pm_1  Higgs cross-section at 14 TeV %%%%%%%%%%%%%%%%
%%%%%%%%%%%%%%%%%%%
\begin{table}
\begin{center}
\renewcommand{\arraystretch}{1.5}
\begin{tabular}{||c||c|c|c|c||}
\hhline{~====}
\multicolumn{1}{c||}{}&\multicolumn{4}{c||}{Cross Section in fb}\\
\hhline{~====}
\multicolumn{1}{c||}{}&BP1&BP2&BP3&BP4\\
\hline\hline
$h_1^\pm h_1^\mp$&148.00&13.00&12.48&166.50\\
\hline
$h_{125} h_1^\pm$&$6.93\times10^{-4}$&$1.82\times10^{-2}$&0.35&0.15\\
\hline
$a_1 h_1^\pm$&$2.14\times10^{-2}$&$2.48\times10^{-3}$&5.12&$9.68\times10^{-7}$\\
\hline
$h_2 h_1^\pm$&0.28&26.42&13.89&334.62\\
\hline
$a_2 h_1^\pm$&292.45&$2.38\times10^{-3}$&12.54&$7.07\times10^{-8}$\\
\hline
$Z h_1^\pm$&$1.44\times10^{-3}$&$3.08\times10^{-2}$&$2.66\times10^{-8}$&0.33\\
\hline
$W^\mp h_1^\pm$&$2.08\times10^{-2}$&0.17&166.21&0.88\\
\hline
$t h_1^\pm$&$8.13\times10^{-2}$&3.50&$4.48\times10^3$&7.60\\
\hline
$t b h_1^\pm$&$3.28\times10^{-2}$&0.21&386.32&3.81\\
\hline
\hline

\end{tabular}
\caption{Production cross-section of the charged Higgs boson $h_1^\pm$ in various channels at the LHC for a center of mass energy of 14 TeV for the three benchmark points. A K-factor of 1.6 has been used.}\label{Hcrosssec}
\end{center}
\end{table}

%%%%%%%%%%%

\section{Final state topologies and simulation at the LHC}\label{sigsim}

The TNMSSM can have a light pseudoscalar $a_1$ with $m_{a_1} \lesssim 60$ GeV (see Table~\ref{bps}). The existence of such light pseudoscalar opens a new mode in the decay of the light charged Higgs boson, \textit{i.e.} $h^\pm_1 \to a_1 W^\pm$, and the light pseudoscalar boson can further decay into $\tau$ or $b$ pairs.  The other possible signature comes from the triplet nature of the light charged Higgs boson, which prompts the $h^\pm_1 \to Z W^\pm$  decay. The chosen benchmark points will focus on these non-standard decays of the charged Higgs boson which are the results of the existence of a singlet and triplet scalar in the spectrum. CalcHEP \cite{calchep}  has been used to generate the decay file SLHA, containing the decay branching ratios and the corresponding mass spectra. The generated events have then been simulated with {\tt PYTHIA} \cite{pythia} via the the SLHA interface \cite{slha}. The simulation at hadronic level has been performed using the {\tt Fastjet-3.0.3} \cite{fastjet} with the {\tt CAMBRIDGE AACHEN} algorithm. We have selected a jet size $R=0.5$ for the jet formation, with the following criteria:

\begin{itemize}
  \item the calorimeter coverage is $\rm |\eta| < 4.5$

  \item the minimum transverse momentum of the jet $ p_{T,min}^{jet} = 10$ GeV and jets are ordered in $p_{T}$
  \item leptons ($\rm \ell=e,~\mu$) are selected with
        $p_T \ge 10$ GeV and $\rm |\eta| \le 2.5$
  \item no jet should be accompanied by a hard lepton in the event
   \item $\Delta R_{lj}\geq 0.4$ and $\Delta R_{ll}\geq 0.2$
  \item Since an efficient identification of the leptons is crucial for our study, we additionally require  
a hadronic activity within a cone of $\Delta R = 0.3$ between two isolated leptons to be $\leq 0.15\, p^{\ell}_T$ GeV, with 
$p^{\ell}_T$ the transverse momentum of the lepton, in the specified cone.

\end{itemize}

%%%%%%%%%%%%%%%%%%%%%%%%%
\begin{figure}[thb]
\begin{center}
\mbox{
\hspace{-1cm}\subfigure[]{
\includegraphics[width=0.56\linewidth]{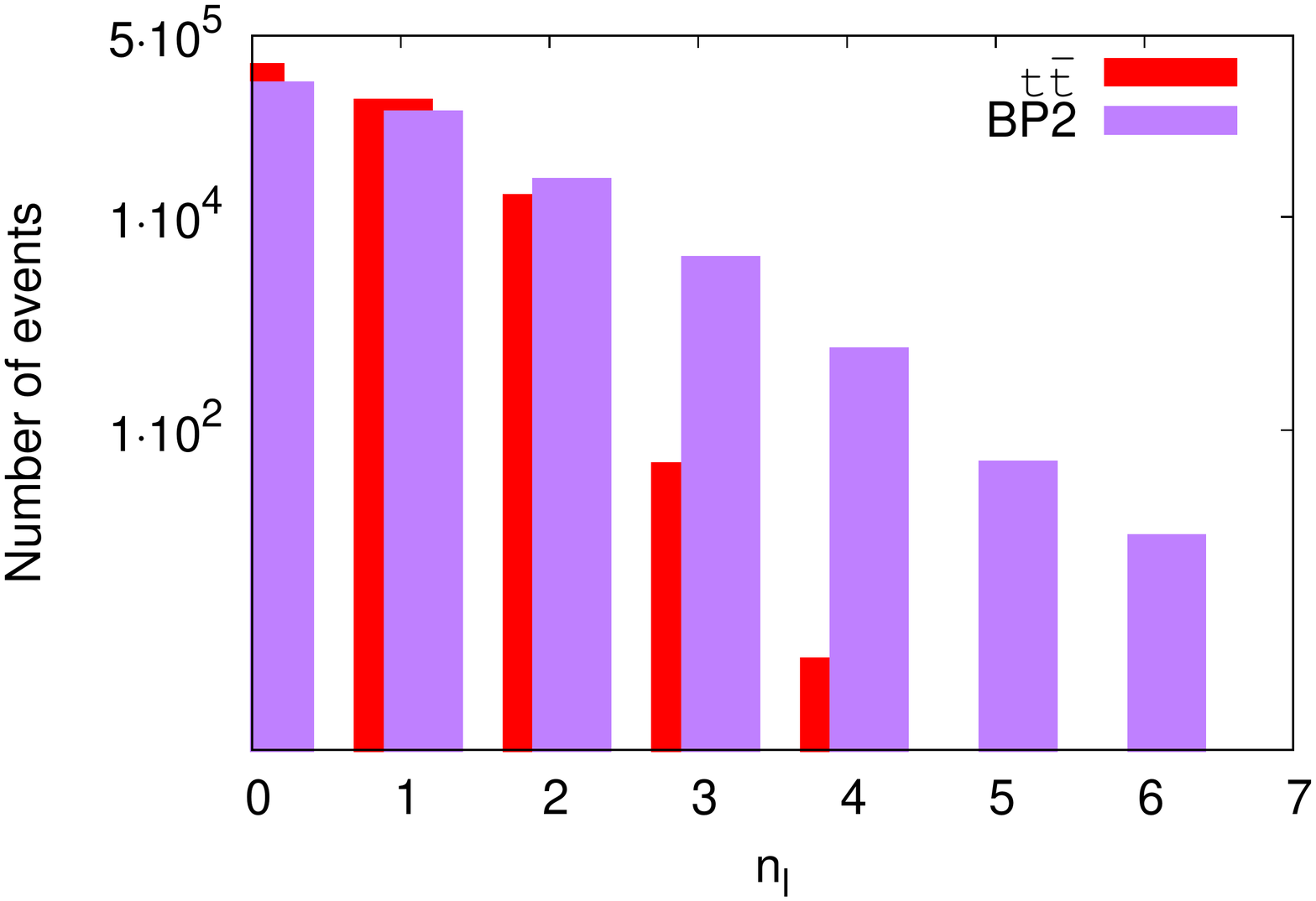}}
\subfigure[]{\includegraphics[width=0.56\linewidth]{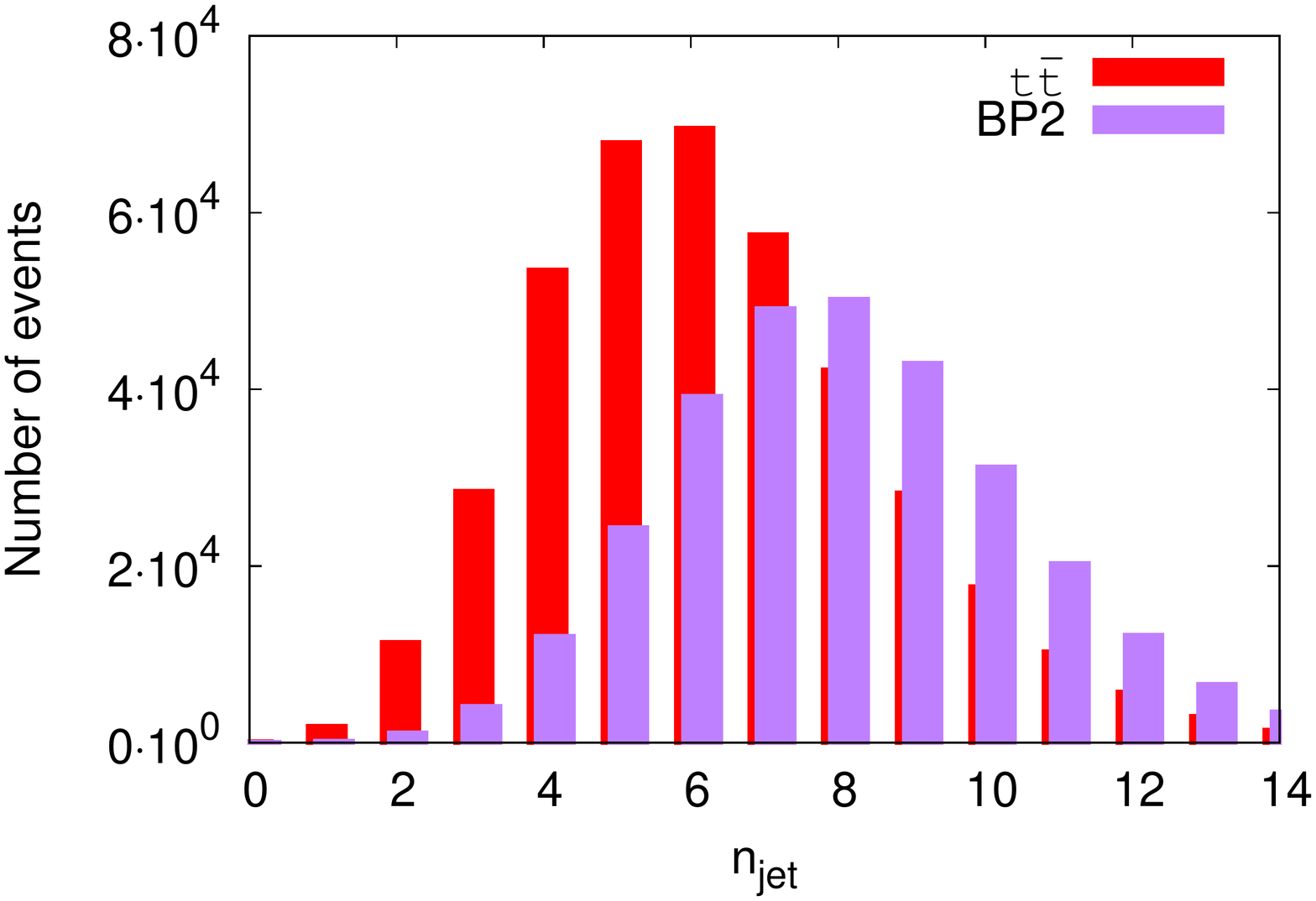}}}
\caption{Jet and lepton multiplicity distributions from BP2 of the $h^\pm_1 h^\mp_1$ signal and the SM background $t\bar{t}$ are shown in (a) and (b) respectively.}\label{lmt}
\end{center}
\end{figure}
%%%%%%%%%%%%%%%%%%%%%%
The non-standard decay products of the light charged Higgs boson $h^\pm_1$ gives rise to gauge bosons $Z$, $W^\pm$ which can further decays into charged leptons. Here we tag $e, \mu$ as charged leptons only and $\tau$ is tagged as jet via its hadronic decay. Figure~\ref{lmt} (a) describes the lepton multiplicity distribution coming from BP2 of the $h^\pm_1 h^\mp_1$ signal and the SM background $t\bar{t}$. Clearly a higher lepton multiplicity is a winner here. Similarly  Figure~\ref{lmt} (b) shows the jet multiplicity distributions  from BP2 of the $h^\pm_1 h^\mp_1$ signal and the SM background $t\bar{t}$. Concerning the signal, these jets are coming from the hadronic decay of $Z, W^\pm, \tau$ as well as the $b$-jets coming from $h^\pm_1$. Similarly to the lepton multiplicity, here also we see that the higher jet-multiplicity is preferred for the signal, which can be used to suppress the SM backgrounds.
%%%%%%%%%%%%%%%%%%%%%%%%%
\begin{figure}[]
\begin{center}
\mbox{
\hspace{-1cm}\subfigure[]{
\includegraphics[width=0.56\linewidth]{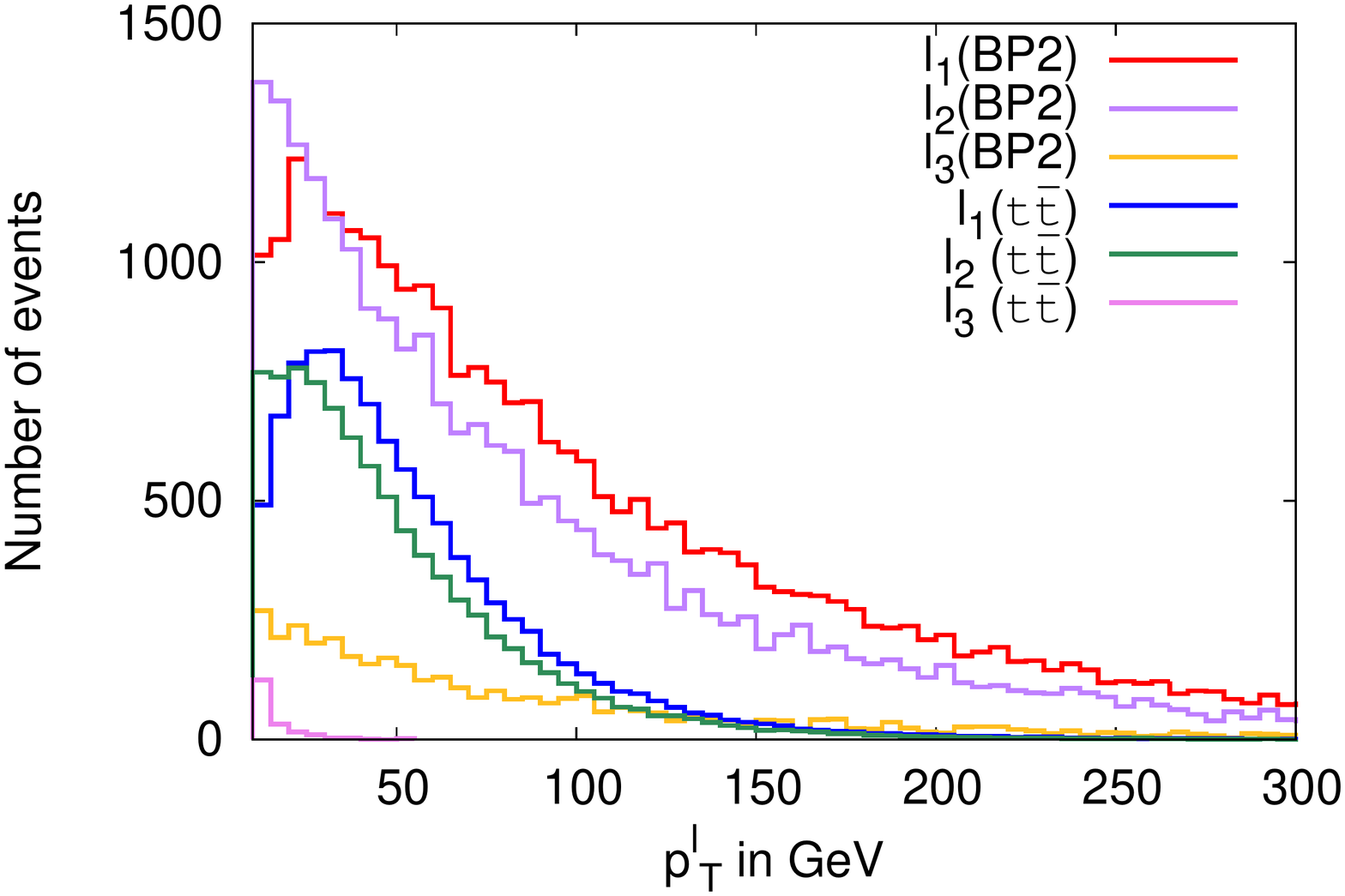}}
\subfigure[]{
\includegraphics[width=0.56\linewidth]{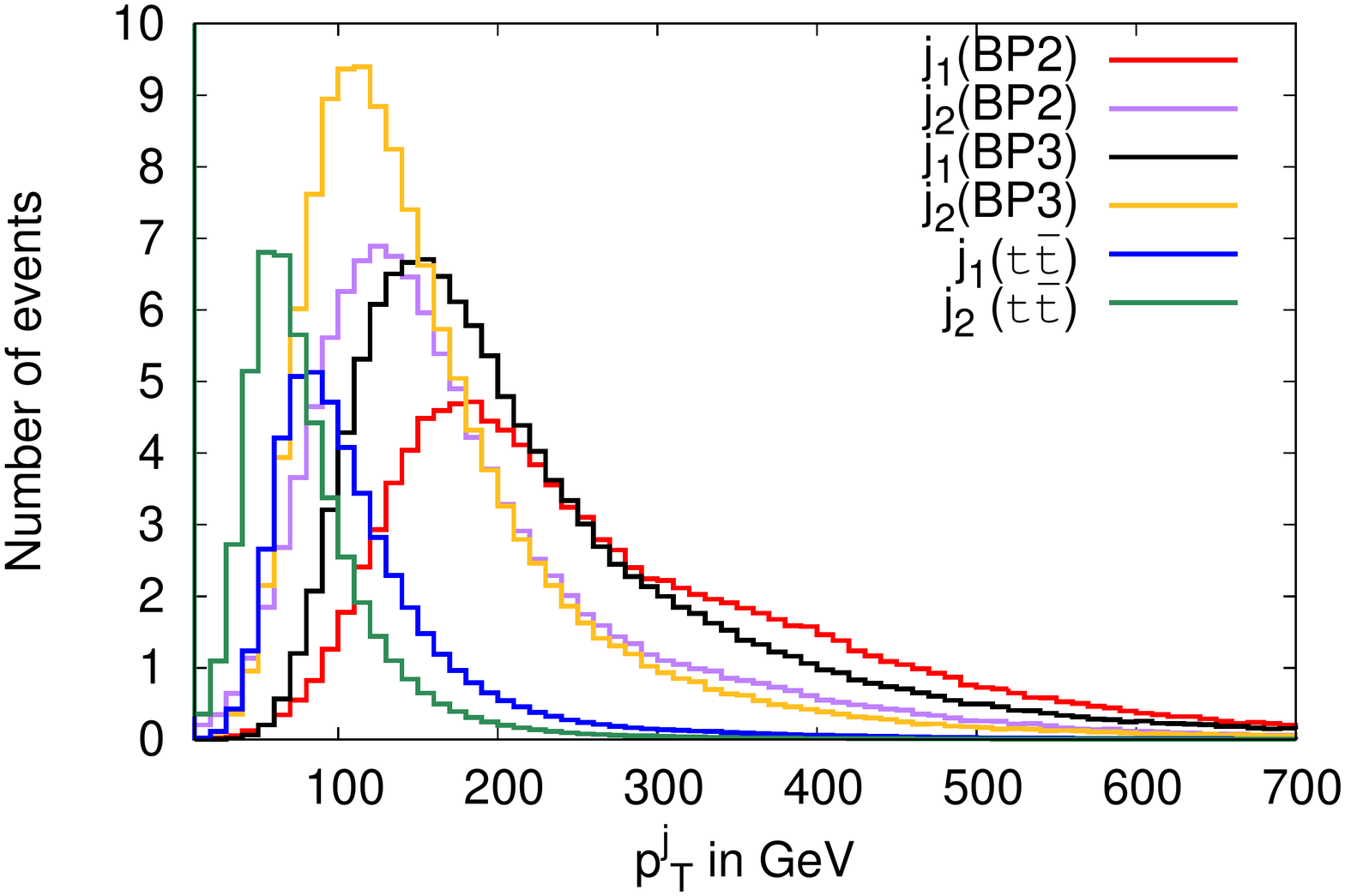}}}
\caption{Lepton and Jet $p_T$ distributions coming from BP2 of the $h^\pm_1 h^\mp_1$ signal and the SM background $t\bar{t}$ are shown in (a) and (b) respectively.}\label{lpt}
\end{center}
\end{figure}
%%%%%%%%%%%%%%%%%%%%%%

We keep the cuts in $p_T$ of the leptons and  the jets relatively low ($p_T \ge 10$ GeV), because they will be generated from the lightest pseudoscalar decays.  Figure~\ref{lpt} (a) shows the lepton $p_T$ distributions for the three leptons in their kinematical order coming from  the signal $h^\pm_1 h^\mp$ for BP2 and from the $t\bar{t}$.  The leptons coming from the signals are of high $p_T$ because they come from a rather heavy charged Higgs boson for BP2 ($m_{h^\pm_1}\sim 340$ GeV), compared to those coming from top quarks. The third lepton coming from the $t\bar{b}$ generally arises from the semi-leptonic $b$ decays and it is rather soft as can be seen from Figure~\ref{lpt} (a).

\begin{figure}[thb]
\begin{center}
\mbox{
\hspace{-1cm}\subfigure[]{
	\includegraphics[width=0.56\linewidth]{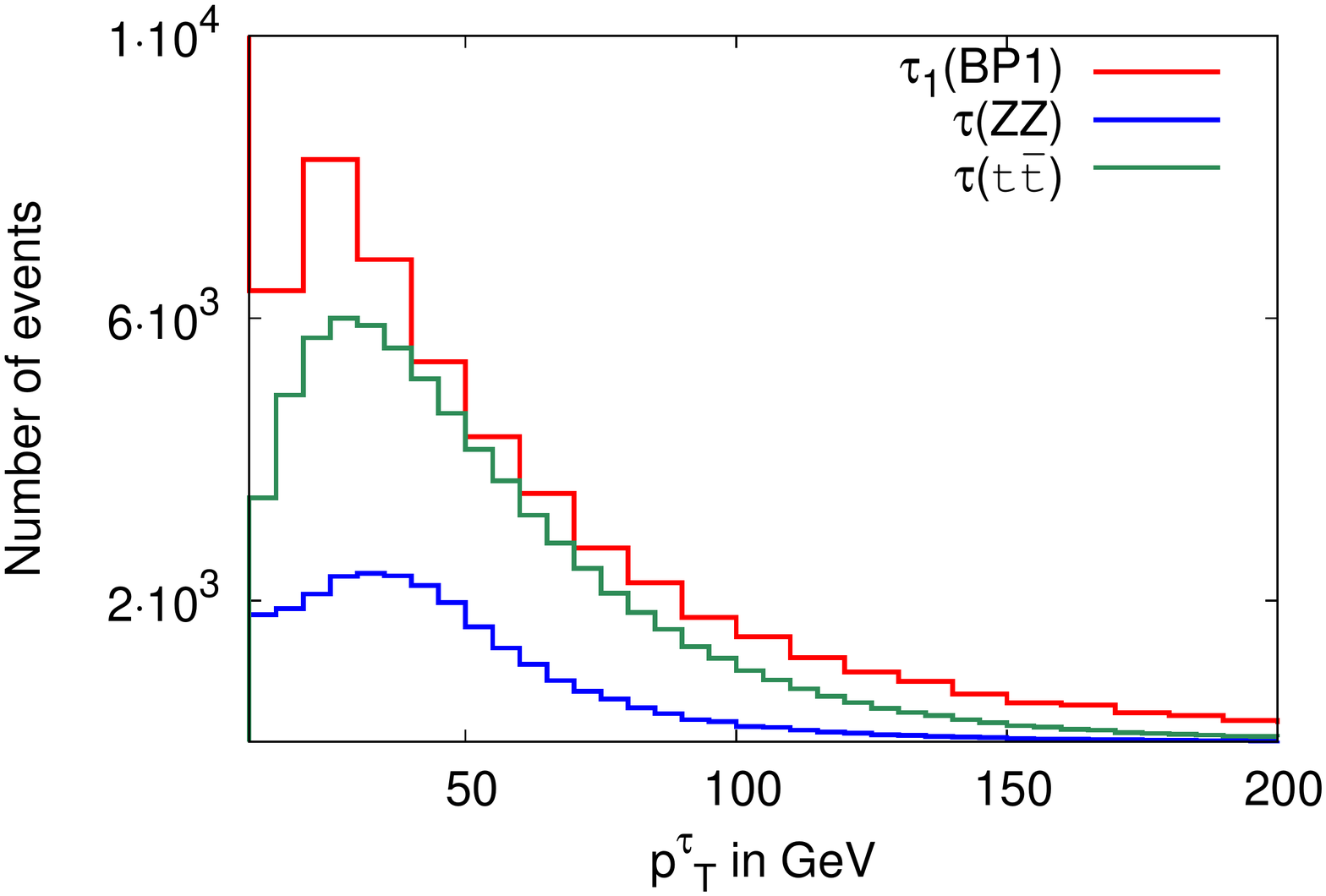}}
	\subfigure[]{\includegraphics[width=0.56\linewidth]{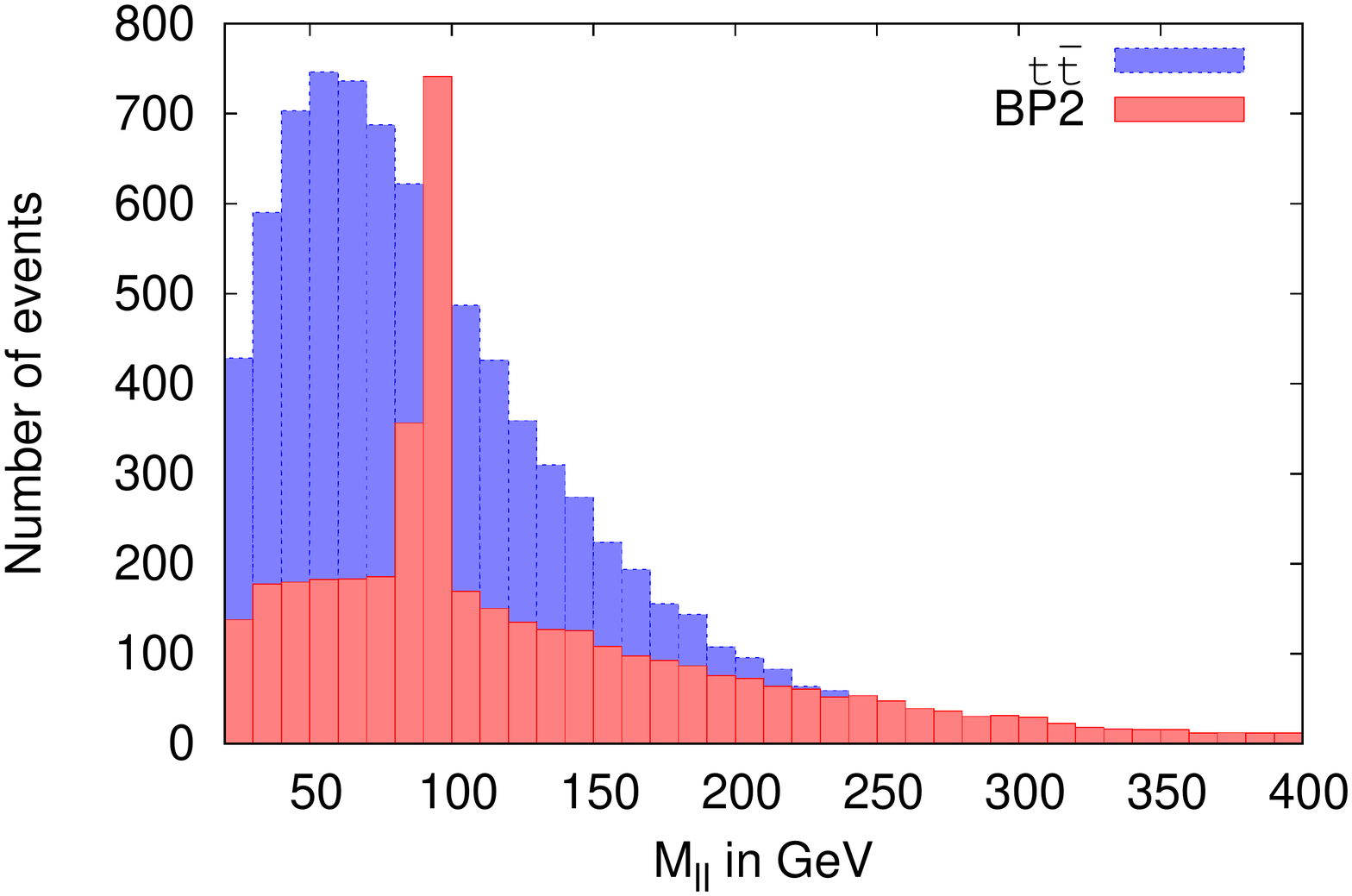}}}
\mbox{
	\hspace{-1cm}\subfigure[]{
\includegraphics[width=0.56\linewidth]{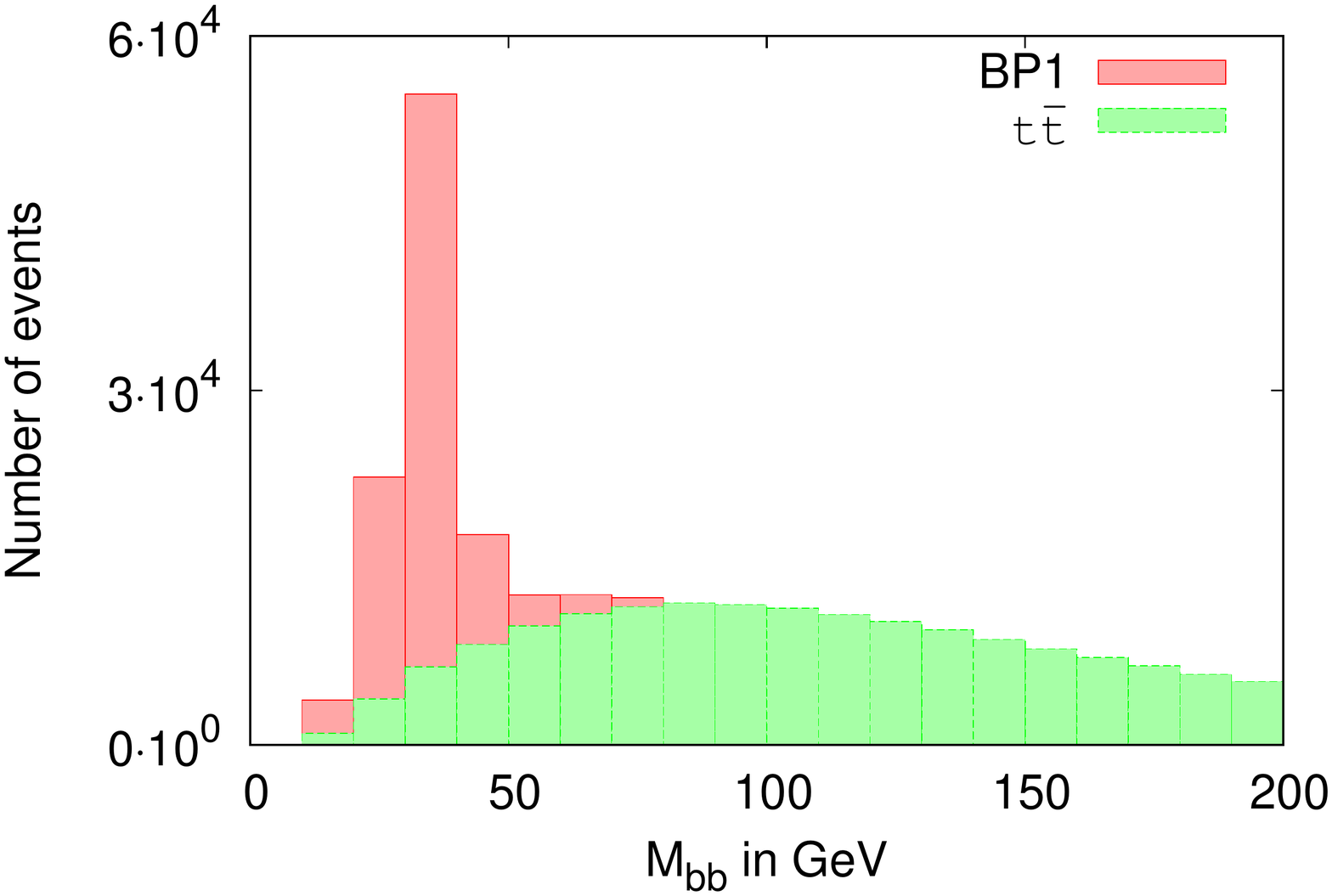}}
\subfigure[]{\includegraphics[width=0.56\linewidth]{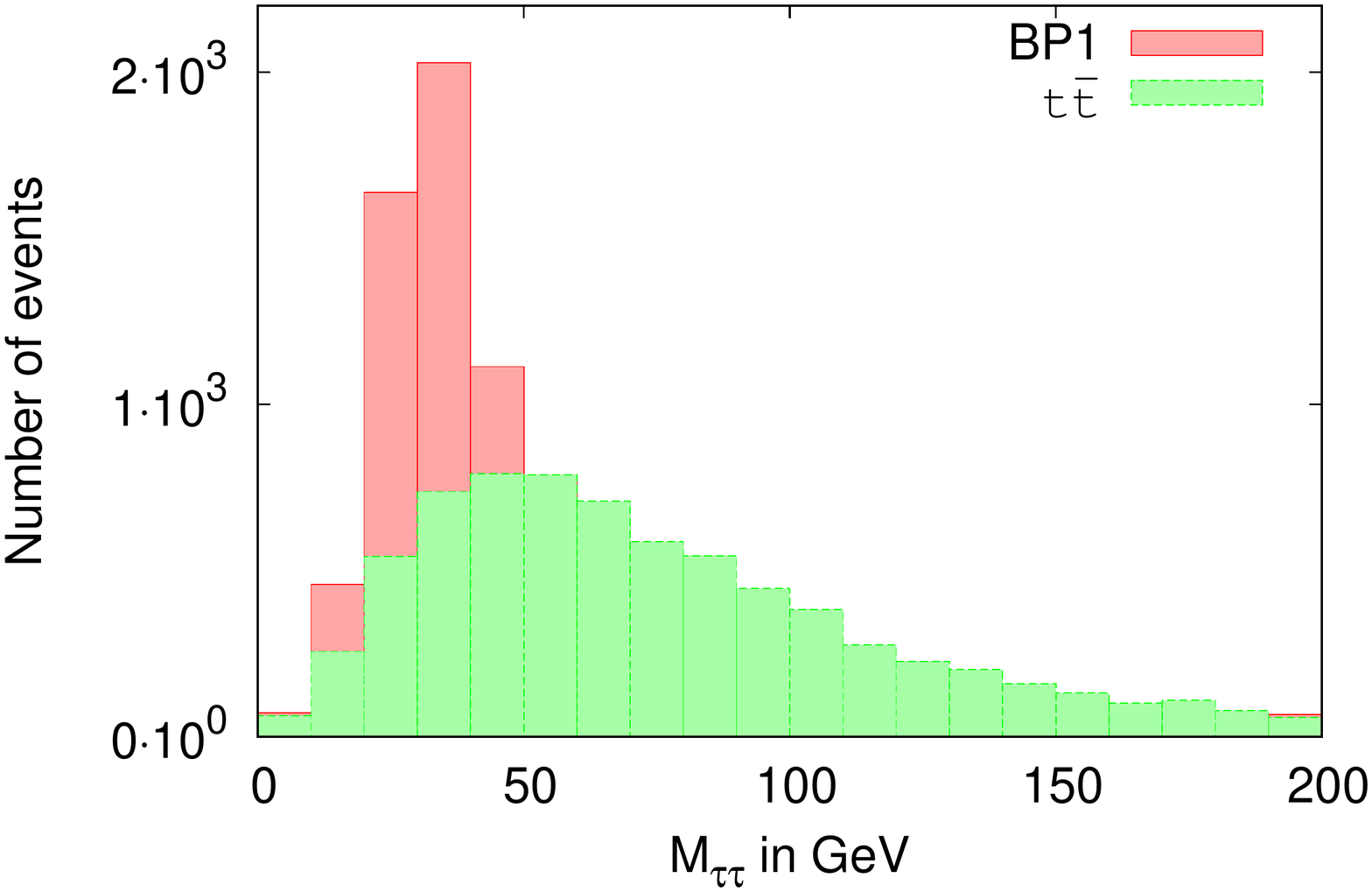}}}
\caption{$\tau$-$p_T$ distributions for $h^\pm_1 h^\mp_1$ signal of BP1, $ZZ$ and $t\bar{t}$ are shown in (a). Invariant mass distribution for the (b) di-lepton, (c) di-$b$-jets,  and (d) di-tau coming from BP2 of the $h^\pm_1 h^\mp_1$ signal, the SM background $t\bar{t}$ and from BP1 of the $h^\pm_1 h^\mp_1$ signal, the SM background $t\bar{t}$ respectively. }\label{invm}
\end{center}
\end{figure}

Figure~\ref{lpt} (b) describes the jet $p_T$ distributions  for the first two $p_T$ ordered jets coming from BP2 and BP3 of the $h^\pm_1 h^\mp_1$ signal and the SM background $t\bar{t}$. We clearly see that a cut $p_T >100$ GeV can reduce such backgrounds
considerably. The tagging efficiency of the jet of the b-quark ($b_{\rm{jet}}$) is obtained through the determination of the secondary vertex  and it is therefore momentum dependent. For this purpose we have taken - for the $b_{\rm{jet}}$'s from $t\bar{t}$ - the single-jet tagging efficiency equal to $0.5$, while for the remaining components of the final state we have followed closely the treatment of \cite{btag}. Here, in the case of the $\tau_{\rm{jet}}$ we have considered the hadronic decay of the $\tau$ to be characterized by at least one charged track with $\Delta R \leq 0.1$ of the candidate $\tau_{\rm{jet}}$ \cite{taujet}. Figure~\ref{invm} (a) shows the $\tau_{\rm{jet}}$ $p_T$ distributions for the BP1 and the dominant SM backgrounds coming from $ZZ$, $t\bar{t}$.
%%%%%%%%%%%%%%%%%%%%%%%%%%%%%%%%%%%%%%%%%%%%%

%%%%%%%%%%%%%%%%%%%%%%%%%
%%%%%%%%%%%%%%%%%%%%%%

%%%%%%%%%%%%%%%%%%%%%%%

Figure~\ref{invm} (b) describes the invariant mass distribution for the di-lepton coming from BP2 of the $h^\pm_1 h^\mp_1$ signal and the SM background $t\bar{t}$. It can be seen that in the case of $t\bar{t}$ both the charged leptons are coming from the corresponding $W^\pm$ decays and hence the invariant mass distribution of lepton pair does not have any peak. However, as the charged Higgs boson decays to $Z,W^\pm$, we see the $Z$ mass peak in the invariant mass distribution. The situation is more interesting in the case where $a_1$ comes from the charged Higgs decay $h^\pm_1 \to a_1 W^\pm$ and then further decays into $b$ or $\tau$ pairs. In this scenario the peak around $m_{a_1}$ is easily visible in the variant mass distribution of  the di-$b$-jets and $\tau$-jet pair for BP1, as shown in Figure~\ref{invm} (c) and Figure \ref{invm} (d) respectively. This happens due to the fact that for  BP1 $\mathcal{B}\textrm{r}(h^\pm_1 \to a_1 W^\pm \sim 97\%$) (See Table~\ref{chdcy2}). Later we will extract these peaks in order to probe such decays modes.

In oder to distinguish the doublet- and triplet-type of charged Higgs boson we have to find final states with very different predictions of number of events and the four benchmark points will present that. We will show that it is also possible to discriminate between the non-standard decay modes of the charged Higgs boson, \textit{i.e.} $h_1^\pm\to a_1W^\pm$ and $h_1^\pm\to ZW^\pm$.

\subsection{$2b+2\tau$}
In BP1, BP2 and BP4, the non-standard decay modes \textit{i.e.} $a_1 W^\pm$ and $ZW^\pm$  of the light charged Higgs bosons are open. This results into the possibility to have $\tau $ and $b$ pairs either from $a_1$ or $Z$ boson. In this subsection we focus our attention on the final state involving $2b + 2\tau$. For BP1 the branching ratio $\mathcal{B}\textrm{r}(h^\pm_1 \to a_1 W^\pm \sim 97\%$), the possibility of having two $a_1$ in the final state is very high. Such $a_1$ produced from the charged Higgs boson decay, mostly decays to $b$ pair ($\sim 95\%$) but also to $\tau$ pairs ($\sim5\%$), \textit{cf}. Table~\ref{a1dcy2}. Thus a suitable choice of the final state is the $2b+2\tau$ one. We expect a scenario where BP1 can be easily probed via this final state and the existence of light pseudoscalar can be also explored. Table~\ref{2b2tau} gives the final state numbers with the cumulative cuts for the benchmark points as well as for the dominant SM backgrounds. For the signal we have included the dominant contributions coming from $h^\pm_1 h^\mp_1$, $h_2 h^\pm_1$ and $a_2 h^\pm_1$, whose cross-sections can be found out in Table~\ref{Hcrosssec}. There are many SM processes with final states involving $b$ coming from top quark or $Z$ boson decays and $\tau$ coming from gauge bosons. The dominant SM backgrounds considered are $t\bar{t}$, $t\bar{t}V$, $tZW$, $VV$ and $VVV$, where $V$ corresponds to $Z$ or $W^\pm$ bosons. A closer look to the final states will tell us that there are $W^\pm$ and $Z$ bosons coming from the charged Higgs and neutral Higgs bosons which can provide additional jets and leptons in the final states. If we tag one of those $W^\pm$ boson via $m_{\rm j\rm j}\sim m_W$ with an additional charged lepton ($e$ or $\mu$), then mostly the $h^\pm_1h^\mp_1$ signal will be filtered over the background.  In that case the signal significance for BP1 jumped to $7.6\sigma$ and such signal can be discovered with an integrated luminosity of 43 fb$^{-1}$ at $5\sigma$ significance. However BP2, BP3 and BP4 will require a much higher luminosity to prove this final state, \textit{i.e.} 3718, 253 and much more than 3000 fb$^{-1}$ respectively. Thus the considered final state can be very effective in the search of the $h^\pm_1 \to a_1 W^\pm$ decay mode, which can potentially discover a light pseudoscalar boson.

%%%%%%%%%%%%%%%%%%%%%%%%%%%%%%%%%%%%%%%
%%%%%%%%%%%%%%%%%%%%%%%%%%%%%%%%%%%%%%
%%%%%%%%%%%%%%%%%% TABLES %%%%%%%%%%%%%%%
%%%%%%%%%%%%%%%%%%%%%%%%%%%%%%%%%%%%%%
%%%%%%%%%%%%%%%%%%%%%%%%%%%%%%%%%%%%%%%

%%%%%%%%%%%%%%%%% 2b 2tau %%%%%%%%%%%%%%%%%%
\begin{table}[]
\begin{center}
\hspace*{-1.50cm}
\renewcommand{\arraystretch}{1.1}
\begin{tabular}{||c|c||c|c|c|c||c|c|c|c||}
\hline\hline
\multicolumn{2}{||c||}{\multirow{2}{*}{Final states}}&\multicolumn{4}{|c||}{Benchmark Points}&\multicolumn{4}{|c||}{Backgrounds}\\
\cline{3-10}
\multicolumn{2}{||c||}{}&BP1&BP2&BP3&BP4&$t\bar t$&$t\bar t\, V$&$t\,Z\,W^\pm$&$VV/VVV$\\
\hline\hline
\multirow{5}{*}{$2b+2\tau$}&$h_1^\pm h_1^\mp$&52.00&4.51&2.71&2.44&\multirow{5}{*}{24834.00}&\multirow{5}{*}{228.14}&\multirow{5}{*}{11.92}&\multirow{5}{*}{590.91}\\
&$h_2 h_1^\pm$&0.48&21.75&12.84&10.60&&&&\\
&$a_2 h_1^\pm$&406.99&0.00&43.57&0.00&&&&\\
&$t h_1^\pm$&0.00&0.00&620.45&0.85&&&&\\
&$tb h_1^\pm$&0.00&0.00&25.84&0.00&&&&\\
\hline
\multirow{5}{*}{$+1\ell+m_{\rm j\rm j}\sim m_W$}&$h_1^\pm h_1^\mp$&10.75&1.02&0.25&0.50&\multirow{5}{*}{0.00}&\multirow{5}{*}{25.36}&\multirow{5}{*}{2.91}&\multirow{5}{*}{0.20}\\
&$h_2 h_1^\pm$&0.11&3.72&2.21&1.34&&&&\\
&$a_2 h_1^\pm$&67.34&0.00&8.98&0.00&&&&\\
&$t h_1^\pm$&0.00&0.00&7.49&0.18&&&&\\
&$tb h_1^\pm$&0.00&0.00&3.47&0.00&&&&\\
\hline\hline
\multicolumn{2}{||c||}{Significance}&$7.60 \sigma$&$0.82 \sigma$&$3.14 \sigma$&$0.37 \sigma$&\multicolumn{4}{|c||}{}\\
%\hline
\multicolumn{2}{||c||}{$\mathcal{L}_{5\sigma}$ (fb$^{-1}$)}&43&3718&253&$\gg3000$&\multicolumn{4}{|c||}{}\\
\hline\hline
\end{tabular}
\caption{The number of events for a $2b+2\tau$ final state at 100 fb$^{-1}$ of luminosity at the LHC with 14 TeV center of mass energy.}\label{2b2tau}
\end{center}
\end{table}
%%%%%%%%%%%%%%%%%%%%%%%%%%%%%%%%%%%%%%%%%%%%%%%%%%%%%%%%

\subsection{$3\ell$}

The dominant modes for a triplet type charged Higgs boson with an existent light pseudoscalar is either $a_1 W^\pm$ or $ZW^\pm$. The final states are, thus, rich in leptons, prompting the multi-leptonic channels. Along with the leptonic final states, there will be $b$-jets coming from the neutral Higgs bosons ($a_1, a_2, h_2$) and $Z$ boson decays. Tagging with two such $b$-jets will help to control the SM backgrounds, but because the production cross-sections of such triplet-like charged Higgs bosons are relatively small, it is possible to probe such final states only at the LHC with higher luminosity. 

We can see in Table~\ref{3l} two different kind of selection for the multi-leptonic final state $3\ell$. In particular $(\ \ )^\dagger$ means $p^{\ell_2}_T \geq 30\, \rm{GeV}$ + $p^{\ell_3}_T \geq 40\, \rm{GeV}$ whereas $(\ \ )^*$ states that no $p_T$ cuts are added. If we consider the final state $(>3\ell+>2j+m_{\ell\ell}\sim m_Z)^\dagger$ then only BP4 can be almost probed at the LHC, having an integrated luminosity for the discovery of 437 fb$^{-1}$. The other benchmark points will require a very high luminosity to probe such final state. In the case of ($>3\ell\,+\,>2b_{\rm{j}})^*$ the required luminosity for the discovery is 567 and 626 fb$^{-1}$ for BP1 and BP3. On the other hand, BP2 and BP4 will require 4216 and 3740 fb$^{-1}$ respectively of luminosity for the discovery which are on the edge of the HL-LHC project \cite{HLLHC}. Finally, if we tag the jets coming from the gauge boson decays, the signal number improve a lot for BP4 and in that case a final state  $(>3\ell+m_{\rm{jj}}\sim m_W)^*$  can probed at the LHC at 14 TeV ECM, having a luminosity for the discovery of 153 fb$^{-1}$. In Table \ref{3l1tau} we present the case where we tag a tau-jet along with  the $\geq 3\ell$ final states. This makes BP1 and BP4 probable with much earlier data, precisely with 48 and 54 fb$^{-1}$ of integrated luminosity respectively. However, such tagging does not help for the other BPs, where the presence of both $a_1$ and $W^\pm$s are less in the final sates.

%%%%%%%%%%%%%%%%% 3l %%%%%%%%%%%%%%%%%%
\begin{table}[thb]
	\begin{center}
		\hspace*{-1.75cm}
		\renewcommand{\arraystretch}{1}
		\begin{tabular}{||c|c||c|c|c|c||c|c|c|c||}
			\hline\hline
			\multicolumn{2}{||c||}{\multirow{2}{*}{Final states}}&\multicolumn{4}{|c||}{Benchmark Points}&\multicolumn{4}{|c||}{Backgrounds}\\
			\cline{3-10}
			\multicolumn{2}{||c||}{}&BP1&BP2&BP3&BP4&$t\bar t$&$t\bar t\, V$&$t\,Z\,W^\pm$&$VV/VVV$\\
			\hline\hline
			\multirow{5}{*}{$(>3\ell)^\dagger$}&$h_1^\pm h_1^\mp$&6.56&8.48&0.06&31.41&\multirow{5}{*}{0.00}&\multirow{5}{*}{348.69}&\multirow{5}{*}{68.14}&\multirow{5}{*}{4355.69}\\
			&$h_2 h_1^\pm$&0.02&21.35&5.80&171.21&&&&\\
			&$a_2 h_1^\pm$&14.64&0.00&2.96&0.00&&&&\\
			&$t h_1^\pm$&0.00&0.00&3.00&2.78&&&&\\
			&$tb h_1^\pm$&0.00&0.00&1.16&0.00&&&&\\
			\hline
			\multirow{5}{*}{$(+>2\rm{j})^\dagger$}&$h_1^\pm h_1^\mp$&6.07&8.32&0.06&30.08&\multirow{5}{*}{0.00}&\multirow{5}{*}{329.52}&\multirow{5}{*}{62.32}&\multirow{5}{*}{2835.27}\\
			&$h_2 h_1^\pm$&0.03&20.64&5.76&162.07&&&&\\
			&$a_2 h_1^\pm$&14.64&0.00&2.96&0.00&&&&\\
			&$t h_1^\pm$&0.00&0.00&3.00&2.66&&&&\\
			&$tb h_1^\pm$&0.00&0.00&1.16&0.00&&&&\\
			\hline
			\multirow{5}{*}{$(+m_{\ell\ell}\sim m_Z)^\dagger$}&$h_1^\pm h_1^\mp$&3.11&5.73&0.01&24.14&\multirow{5}{*}{0.00}&\multirow{5}{*}{236.40}&\multirow{5}{*}{59.12}&\multirow{5}{*}{2477.44}\\
			&$h_2 h_1^\pm$&0.00&12.97&4.63&102.39&&&&\\
			&$a_2 h_1^\pm$&2.93&0.00&1.71&0.00&&&&\\
			&$t h_1^\pm$&0.00&0.00&0.00&2.29&&&&\\
			&$tb h_1^\pm$&0.00&0.00&0.39&0.00&&&&\\
			\hline\hline
			\multicolumn{2}{||c||}{Significance}&$0.11 \sigma$&$0.35 \sigma$&$0.13 \sigma$&$2.39 \sigma$&\multicolumn{4}{|c||}{}\\
			%\cline{1-5}
			\multicolumn{2}{||c||}{$\mathcal{L}_{5\sigma}$ (fb$^{-1}$)}&$\gg3000$&$\gg3000$&$\gg3000$&437&\multicolumn{4}{|c||}{}\\
			\hline\hline
			\multirow{5}{*}{($>3\ell\,+\,>2b_{\rm{j}})^*$}&$h_1^\pm h_1^\mp$&6.07&3.33&0.12&2.83&\multirow{5}{*}{0.00}&\multirow{5}{*}{294.55}&\multirow{5}{*}{17.74}&\multirow{5}{*}{33.01}\\
			&$h_2 h_1^\pm$&0.04&11.13&6.94&11.38&&&&\\
			&$a_2 h_1^\pm$&35.14&0.00&4.81&0.00&&&&\\
			&$t h_1^\pm$&0.00&0.00&25.00&1.32&&&&\\
			&$tb h_1^\pm$&0.00&0.00&2.31&0.00&&&&\\
			\hline\hline
			\multicolumn{2}{||c||}{Significance}&$2.10 \sigma$&$0.77 \sigma$&$2.00 \sigma$&$0.82 \sigma$&\multicolumn{4}{|c||}{}\\
			%\cline{1-5}
			\multicolumn{2}{||c||}{$\mathcal{L}_{5\sigma}$ (fb$^{-1}$)}&567&4216&626&3740&\multicolumn{4}{|c||}{}\\
			\hline\hline
			\multirow{5}{*}{$(>3\ell+m_{\rm{jj}}\sim m_W)^*$}&$h_1^\pm h_1^\mp$&6.91&7.73&0.02&45.90&\multirow{5}{*}{127.55}&\multirow{5}{*}{270.12}&\multirow{5}{*}{63.49}&\multirow{5}{*}{2804.87}\\
			&$h_2 h_1^\pm$&0.02&16.37&5.87&187.50&&&&\\
			&$a_2 h_1^\pm$&17.57&0.00&3.33&0.00&&&&\\
			&$t h_1^\pm$&0.00&0.00&5.99&6.26&&&&\\
			&$tb h_1^\pm$&0.00&0.00&0.00&0.00&&&&\\
			\hline\hline
			\multicolumn{2}{||c||}{Significance}&$0.43 \sigma$&$0.42 \sigma$&$0.27 \sigma$&$4.05 \sigma$&\multicolumn{4}{|c||}{}\\
			%\cline{1-5}
			\multicolumn{2}{||c||}{$\mathcal{L}_{5\sigma}$ (fb$^{-1}$)}&$\gg3000$&$\gg3000$&$\gg3000$&153&\multicolumn{4}{|c||}{}\\
			\hline\hline
		\end{tabular}
		\caption{The number of events for a $>3\ell$ final state at 100 fb$^{-1}$ of luminosity at the LHC with 14 TeV center of mass energy. Here $(\ \ )^\dagger$ means $p^{\ell_2}_T \geq 30\, \rm{GeV}$ + $p^{\ell_3}_T \geq 40\, \rm{GeV}$
			whereas $(\ \ )^*$ states that no $p_T$ cuts are added.}\label{3l}
	\end{center}
\end{table}
%%%%%%%%%%%%%%%%%%%%%%%%%%%%%%%%%%%%%%%%%%%%%

%%%%%%%%%%%%%%%%% 3l 1tau %%%%%%%%%%%%%%%%%%
\begin{table}[h]
	\begin{center}
		%\hspace*{-1.0cm}
		\renewcommand{\arraystretch}{1.2}
		\begin{tabular}{||c|c||c|c|c|c||c|c|c|c||}
			\hline\hline
			\multicolumn{2}{||c||}{\multirow{2}{*}{Final states}}&\multicolumn{4}{|c||}{Benchmark Points}&\multicolumn{4}{|c||}{Backgrounds}\\
			\cline{3-10}
			\multicolumn{2}{||c||}{}&BP1&BP2&BP3&BP4&$t\bar t$&$t\bar t\, V$&$t\,Z\,W^\pm$&$VV/VVV$\\
			\hline\hline
			\multirow{5}{*}{$3\ell+ 1\tau$}&$h_1^\pm h_1^\mp$&95.71&19.59&0.25&83.25&\multirow{5}{*}{196.24}&\multirow{5}{*}{263.58}&\multirow{5}{*}{75.60}&\multirow{5}{*}{5009.14}\\
			&$h_2 h_1^\pm$&0.24&38.05&7.08&439.47&&&&\\
			&$a_2 h_1^\pm$&468.48&0.00&14.35&0.00&&&&\\
			&$t h_1^\pm$&0.00&0.00&66.67&7.83&&&&\\
			&$tb h_1^\pm$&0.00&0.00&23.14&0.00&&&&\\
			\hline\hline
			\multicolumn{2}{||c||}{Significance}&$7.22 \sigma$&$0.77 \sigma$&$1.48 \sigma$&$6.81 \sigma$&\multicolumn{4}{|c||}{}\\
			%\cline{1-5}
			\multicolumn{2}{||c||}{$\mathcal{L}_{5\sigma}$ (fb$^{-1}$)}&48&4216&$1138$&54&\multicolumn{4}{|c||}{}\\
			\hline\hline
		\end{tabular}
		\caption{The number of events for a $3\ell+1\tau$ final state at 100 fb$^{-1}$ of luminosity at the LHC with 14 TeV center of mass energy.}\label{3l1tau}
	\end{center}
\end{table}
%%%%%%%%%%%%%%%%%%%%%%%%%%%%%%%%%%%%%%%%%%%%%

%%%%%%%%%%%%%%%%%%%%%%%%%%%%%%%%%%%%%%%
%%%%%%%%%%%%%%%%%%%%%%%%%%%%%%%%%%%%%%
%%%%%%%%%%%%%%%%%%%%%%%%%%%%%%%%%%%%%%%
%%%%%%%%%%%%%%%%%%%%%%%%%%%%%%%%%%%%%%
%%%%%%%%%%%%%%%%%%%%%%%%%%%%%%%%%%%%%%%

\subsection{$3\tau$ }

In this subsection we consider the case where the $\tau$'s are coming from the pseudoscalars  and $Z$ boson decay into $\tau$ pairs or from the $W^\pm$ boson decay. The multi-tau final state, with $\geq 3\tau$, is mostly background free. Moreover, if we tag such final state with a charged lepton which arises from the $W^\pm$ decay, we reduce more efficiently the residual SM background. The dominant decay mode for BP1 and BP4 is $h^\pm_1 \to a_1 W^\pm$ and $h_1^\pm\to Z W^\pm$ respectively whereas in the case of BP2 both modes compete, making such final state viable. However, in the case of BP3, such $\tau$ states can only appear when the charged Higgs boson is produced in association of $a_2$ or $h_2$.
 
We list the number of the events for $\geq 3\tau +\geq 1\ell$ final state for the benchmark points and the dominant SM backgrounds in Table~\ref{3tau}. We see that BP1 reach the luminosity for the discovery at 71 fb$^{-1}$. For the other benchmark points it requires higher luminosities, in particular much more than 3000 fb$^{-1}$ for BP2. The final state $\geq 3\tau +\geq 1\ell$ became a possible discovery mode for BP3 and BP4 in the case of HL-LHC.

%%%%%%%%%%%%%%%%% 3tau %%%%%%%%%%%%%%%%%%
\begin{table}[h]
\begin{center}
%\hspace*{-1.0cm}
\renewcommand{\arraystretch}{1.1}
\begin{tabular}{||c|c||c|c|c|c||c|c|c|c||}
\hline\hline
\multicolumn{2}{||c||}{\multirow{2}{*}{Final states}}&\multicolumn{4}{|c||}{Benchmark Points}&\multicolumn{4}{|c||}{Backgrounds}\\
\cline{3-10}
\multicolumn{2}{||c||}{}&BP1&BP2&BP3&BP4&$t\bar t$&$t\bar t\, V$&$t\,Z\,W^\pm$&$VV/VVV$\\
\hline\hline
\multirow{5}{*}{$\geq3\tau$}&$h_1^\pm h_1^\mp$&26.69&1.62&0.20&4.22&\multirow{5}{*}{0.00}&\multirow{5}{*}{23.42}&\multirow{5}{*}{4.94}&\multirow{5}{*}{964,12}\\
&$h_2 h_1^\pm$&0.11&5.38&1.78&19.85&&&&\\
&$a_2 h_1^\pm$&204.96&0.00&8.52&0.00&&&&\\
&$t h_1^\pm$&0.00&0.00&10.00&0.47&&&&\\
&$tb h_1^\pm$&0.00&0.00&1.93&0.00&&&&\\
\hline
\multirow{5}{*}{$+\geq1\ell$}&$h_1^\pm h_1^\mp$&7.20&0.48&0.04&1.00&\multirow{5}{*}{0.00}&\multirow{5}{*}{5.19}&\multirow{5}{*}{0.77}&\multirow{5}{*}{68.67}\\
&$h_2 h_1^\pm$&0.03&1.55&0.30&6.36&&&&\\
&$a_2 h_1^\pm$&64.42&0.00&1.94&0.00&&&&\\
&$t h_1^\pm$&0.00&0.00&5.00&0.00&&&&\\
&$tb h_1^\pm$&0.00&0.00&0.77&0.00&&&&\\
\hline\hline
\multicolumn{2}{||c||}{Significance}&$5.92 \sigma$&$0.19 \sigma$&$0.88 \sigma$&$0.81\sigma$&\multicolumn{4}{|c||}{}\\
%\cline{1-5}
\multicolumn{2}{||c||}{$\mathcal{L}_{5\sigma}$ (fb$^{-1}$)}&71&$\gg3000$&$3190$&3784&\multicolumn{4}{|c||}{}\\
\hline\hline
\end{tabular}
\caption{The number of events for a $>3\tau$ final state at 100 fb$^{-1}$ of luminosity at the LHC with 14 TeV center of mass energy.}\label{3tau}
\end{center}
\end{table}
%%%%%%%%%%%%%%%%%%%%%%%%%%%%%%%%%%%%%%%%%%%%%

%%%%%%%%%%%%%%%%% 4l %%%%%%%%%%%%%%%%%%
\begin{table}[h]
\begin{center}
%\hspace*{-1.0cm}
\renewcommand{\arraystretch}{1.1}
\begin{tabular}{||c|c||c|c|c|c||c|c|c|c||}
\hline\hline
\multicolumn{2}{||c||}{\multirow{2}{*}{Final states}}&\multicolumn{4}{|c||}{Benchmark Points}&\multicolumn{4}{|c||}{Backgrounds}\\
\cline{3-10}
\multicolumn{2}{||c||}{}&BP1&BP2&BP3&BP4&$t\bar t$&$t\bar t\, V$&$t\,Z\,W^\pm$&$VV/VVV$\\
\hline\hline
\multirow{5}{*}{\shortstack{$\geq4\ell$\\+$p^{\ell_1}_T \geq 50\, \rm{GeV}$\\+$p^{\ell_2}_T \geq 40\, \rm{GeV}$}}&$h_1^\pm h_1^\mp$&14.80&11.27&0.12&44.96&\multirow{5}{*}{0.00}&\multirow{5}{*}{215.85}&\multirow{5}{*}{59.13}&\multirow{5}{*}{2423.16}\\
&$h_2 h_1^\pm$&0.02&27.12&2.59&255.43&&&&\\
&$a_2 h_1^\pm$&29.28&0.00&3.24&0.00&&&&\\
&$t h_1^\pm$&0.00&0.00&16.67&3.50&&&&\\
&$tb h_1^\pm$&0.00&0.00&3.86&0.00&&&&\\
\hline\hline
\multicolumn{2}{||c||}{Significance}&$0.84 \sigma$&$0.73 \sigma$&$0.51 \sigma$&$5.55 \sigma$&\multicolumn{4}{|c||}{}\\
%\cline{1-5}
\multicolumn{2}{||c||}{$\mathcal{L}_{5\sigma}$ (fb$^{-1}$)}&3543&4691&$\gg3000$&81&\multicolumn{4}{|c||}{}\\
\hline\hline
\end{tabular}
\caption{The number of events for a $4\ell$ final state at 100 fb$^{-1}$ of luminosity at the LHC with 14 TeV center of mass energy. The $p_T$ of the first lepton is greater than 50 GeV whereas the $p_T$ of the second lepton is greater than 40 GeV. No selections are applied on the other leptons.}\label{4l}
\end{center}
\end{table}
%%%%%%%%%%%%%%%%%%%%%%%%%%%%%%%%%%%%%%%%%%%%%

\subsection{$4\ell$}
Finally we have considered the multi-leptonic final state with $\geq4\ell$. The dominant backgrounds are $VV$ and $VVV$, where $V=Z,W^\pm$. We report in Table \ref{4l} the number of events for the signal and the dominant backgrounds. We can see that only BP4 reach a $5\sigma$ of discovery within the luminosity of LHC, precisely at 81 fb$^{-1}$. Such a discovery can be achieved at the HL-LHC in the case of BP1. 

\section{Reconstruction of the charged Higgs boson}\label{recon}
\begin{figure}
	\begin{center}
		\mbox{
			\hspace{-1cm}\subfigure[]{
				\includegraphics[width=0.56\linewidth]{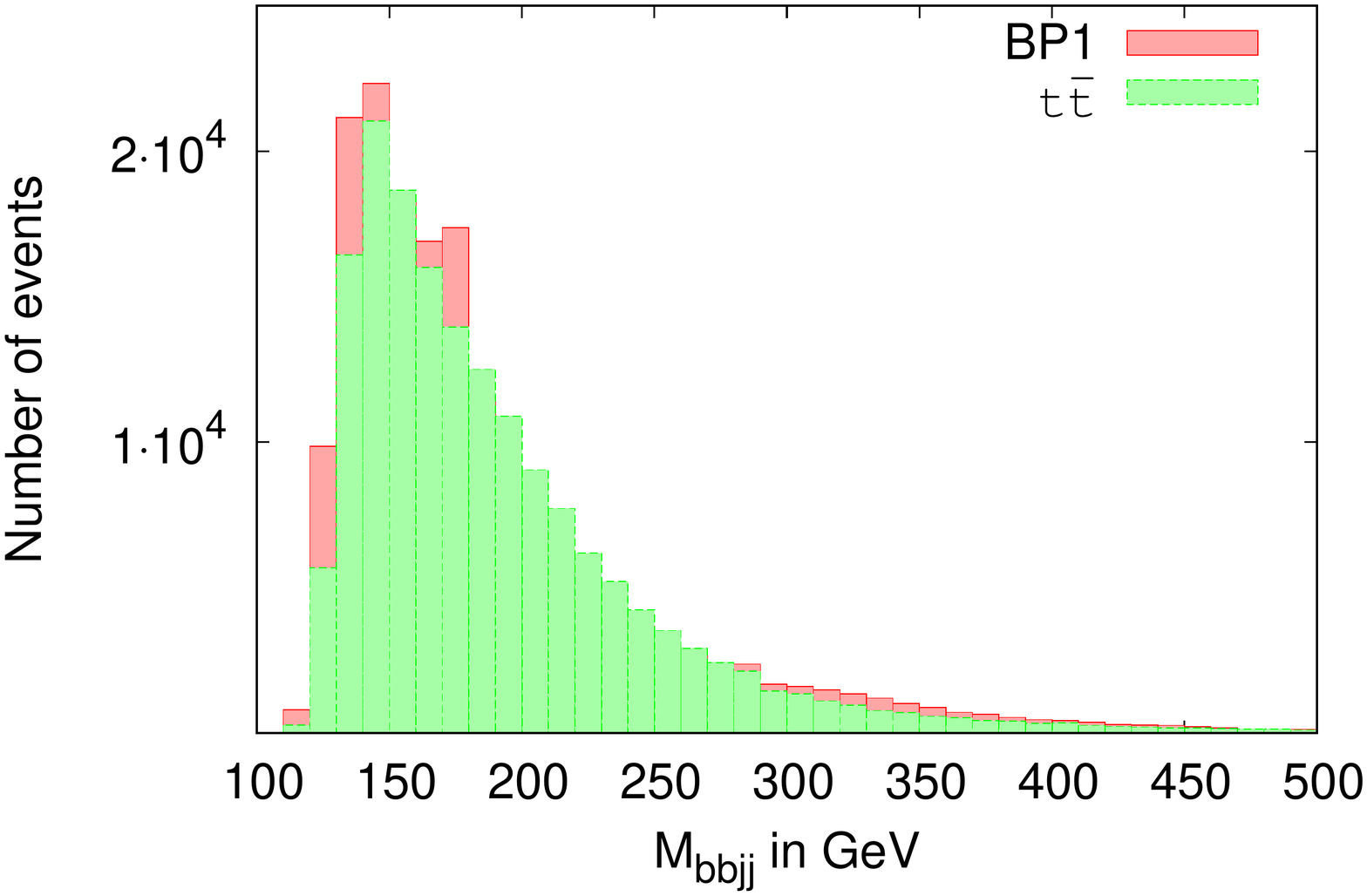}}
			\subfigure[]{\includegraphics[width=0.56\linewidth]{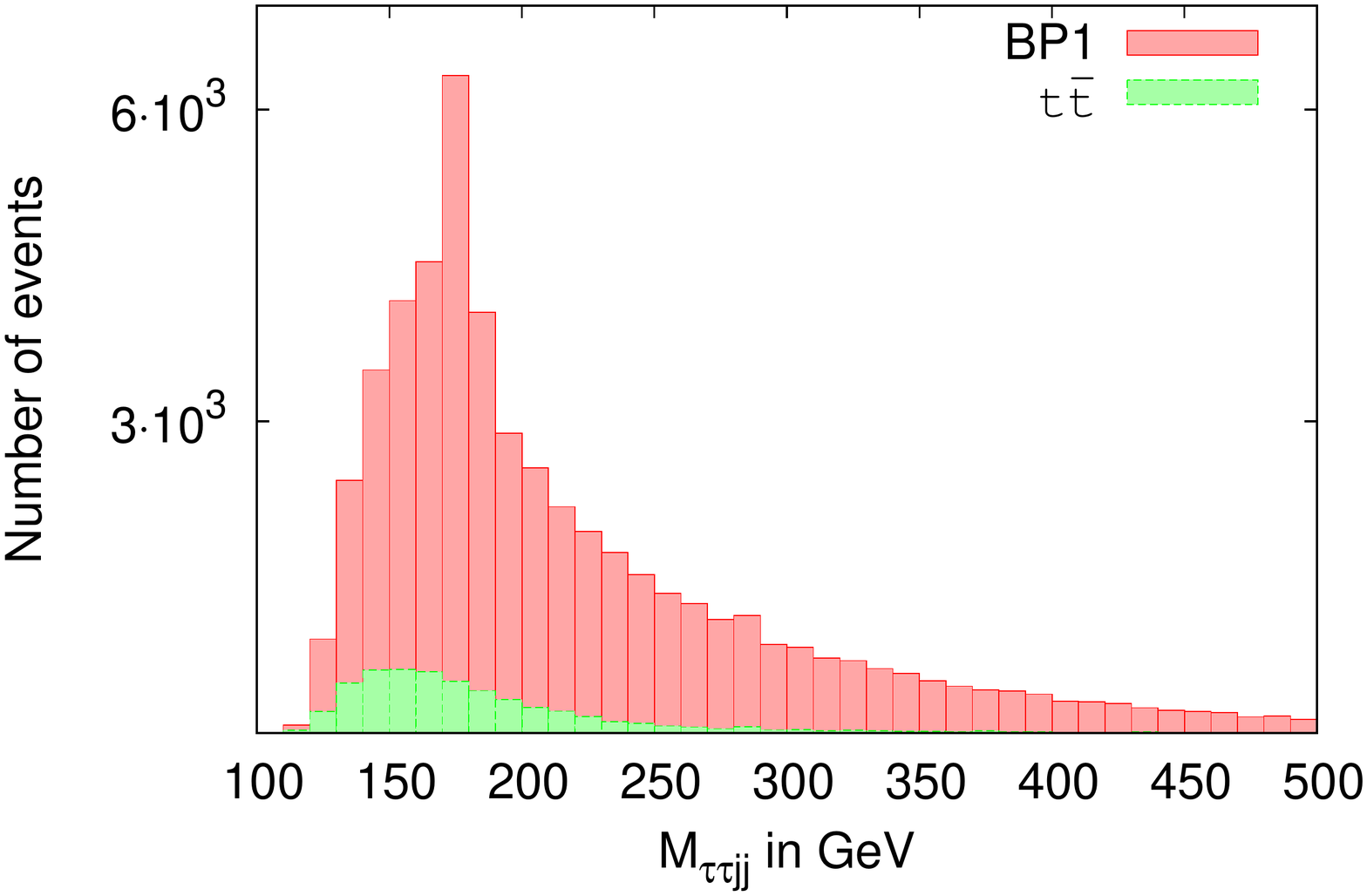}}}
		\caption{Invariant mass distributions for the   reconstruction of the charged Higgs boson mass for the benchmark points. (a) describes  $bbjj$ invariant mass distributions, where $b$-jet pairs are coming from $a_1$ peak and normal jet pairs are coming from $W^\pm$ peak. (b) shows $\tau\tau jj$ invariant mass distribution where $\tau$-jet pairs are coming from $a_1$ mass peak and the normal jet-pairs are coming from $W^\pm$ mass peak respectively.}\label{inviijj}
	\end{center}
\end{figure}

In this section we concentrate on the final states appropriate to the respective decay channels which can be used to reconstruct the charged Higgs boson mass, $m_{h^\pm_1}$. We can see from Table~\ref{bps} and Table~\ref{chdcy2} that for BP1
$h^\pm_1 \to a_1 W^\pm$ and for BP4 $h^\pm \to Z W^\pm$ are the most dominant decay modes, signifying the existence of light pseudoscalar and the triplet nature of the charged Higgs boson respectively.  The light-pseudoscalar mass can be reconstructed via the invariant mass of $b$-jet or $\tau$-jet pairs as can be seen from Figure~\ref{invm} (c) and Figure~\ref{invm} (d) respectively. 
For this purpose we first reconstruct $a_1$ mass peak via $b\bar{b}$ decay mode and we consider $b$-jets within $\pm 7.5 $ GeV of the $b$-jet invariant mass distribution around the $a_1$ mass peak for the reconstruction of the charged Higgs boson. Similarly we reconstruct $W^\pm$ mass peak from di-jet invariant mass distribution. $b$-jets from the selected window around $a_1$ peak are taken along with the jets within the $W^\pm$ mass window for the distribution of $2b2j$. Figure~\ref{inviijj} (a) shows such distributions for the benchmark points along with the dominant SM backgrounds. It is clearly visible that for BP1, we can reconstruct that charged Higgs boson. 
%%%%%%%%%%%%%%%%%%%%%%%%%
%%%%%%%%%%%%%%%%%%%%%%

Next we move to $2\tau2j$ invariant mass distribution as shown in Figure~\ref{inviijj} (b). Here we have consider the $\tau$-jets coming from the $\pm 10$ GeV window of the $a_1$ mass peak. The normal jet pairs are taken from $\pm 10$ GeV of the $W^\pm$ mass window. Reconstruction via both $2b2j$ and $2\tau2j$ are possible but the discovery reach requires very high luminosity run at the LHC. For the current LHC run, thus we investigate the final states in much cleaver way probing only the $a_1$ mass peak while dealing with $b$ and $\tau$-jets. We see that for $h^\pm_1 \to Z W^\pm \to 2l 2j$ mode, the charged Higgs mass reconstruction is possible with much earlier data due to clean and easy tagging of the charged leptons. 

In Table~\ref{bbjja} we present the final state numbers for the benchmark points and the dominant SM backgrounds to probe $h^\pm \to a_1 W^\pm$ decay modes. The significance (required luminosity for $5\sigma$ discovery) for the benchmark points are given in the last column. The final state is comprised of  $\geq 2\tau-\rm{jets} + \geq 2b-\rm{jets} + \geq 1 \ell$, where the $\tau$s and $b$s are coming from the light pseudoscalar and one charged lepton can from one of the two $W^\pm$ bosons coming from charged Higgs pair. After that selection events within $\pm 10$ GeV of $a_1$ mass peak via $b$-jet pair invariant mass have been selected and presented in Table~\ref{bbjja}  at an  integrated luminosity 100 fb$^{-1}$. It can be seen that only BP1 for which the dominant decay mode is $h^\pm \to a_1 W^\pm$, a discovery of $5.37\sigma$ can be achieved at 100 fb$^{-1}$ of integrated luminosity. For other benchmark points one requires very high luminosities as can  be read from Table~\ref{bbjja}.
%%%%%%%%%%%%%%%%%%%%%%%%%%%%%%%%%%%%%%%%%%
\begin{table}[h]
	\begin{center}
		%\hspace*{-1.0cm}
		\renewcommand{\arraystretch}{1.3}
		\begin{tabular}{||c||c|c|c|c|c||c|c|c|c||c||}
			\hline\hline
			\multirow{2}{*}{}&\multicolumn{5}{|c||}{Signals}&\multicolumn{4}{|c||}{Backgrounds}&Significance\\
			\cline{2-10}
			&$h_1^\pm h_1^\mp$&$h_2 h_1^\pm$&$a_2 h_1^\pm$&$t h_1^\pm$&$tb h_1^\pm$&$t\bar t$&$t\bar t\, V$&$t\,Z\,W^\pm$&$VV/VVV$&$\mathcal{L}_{5\sigma}$ (fb$^{-1}$)\\
			\hline\hline
			BP1&6.81&0.04&23.42&0.00&0.00&0.00&1.41&0.01&0.10&$5.37 \sigma$ (87)\\
			BP2&0.07&0.34&0.00&0.00&0.00&0.00&1.57&0.01&0.10&$0.28 \sigma$ ($\gg3000$)\\
			BP3&0.01&0.19&0.56&0.00&0.39&0.00&1.20&0.01&0.00&$0.75 \sigma$ (4471)\\
			BP4&0.17&0.45&0.00&0.02&0.00&0.00&2.15&0.01&0.00&$0.40 \sigma$ ($\gg3000$)\\
			\hline\hline
		\end{tabular}
\caption{The number of event combination of $b$-jet pair at 100 fb$^{-1}$ integrated luminosity, where $b$-jets are within $\pm 10 $ GeV of  $a_1$ mass peak in a final state comprised of $\geq 2\tau-\rm{jets} + \geq 2b-\rm{jets} + \geq 1 \ell$. The significance (required luminosity for $5\sigma$ discovery) for the benchmark points are given in the last column. }\label{bbjja}
\end{center}
\end{table}

Similarly Table~\ref{tautaujja} presents the final states comprised of $\geq 2\tau-\rm{jets} + \geq 2b-\rm{jets} + \geq 1 \ell$, where the $\tau$s and $b$s are coming from the light pseudoscalar and one charged lepton can from one of the two $W^\pm$ bosons coming from charged Higgs pair. However, in this case we select the events for which the $\tau$-jet pairs falls within $\pm 10$ GeV of $a_1$ mass peak. The significance at an integrated luminosity of 100 fb$^{-1}$ (required luminosity for $5 \sigma$ discovery) for the benchmark points are given in the last column of Table~\ref{tautaujja}. In this case also BP1 reaches a signal significance of $6.73\sigma$ whereas for other benchmark points a discovery of $5\sigma$ requires higher luminosities. 

\begin{table}[h]
	\begin{center}
		%\hspace*{-1.0cm}
		\renewcommand{\arraystretch}{1.3}
		\begin{tabular}{||c||c|c|c|c|c||c|c|c|c||c||}
			\hline\hline
			\multirow{2}{*}{}&\multicolumn{5}{|c||}{Signals}&\multicolumn{4}{|c||}{Backgrounds}&Significance\\
			\cline{2-10}
			&$h_1^\pm h_1^\mp$&$h_2 h_1^\pm$&$a_2 h_1^\pm$&$t h_1^\pm$&$tb h_1^\pm$&$t\bar t$&$t\bar t\, V$&$t\,Z\,W^\pm$&$VV/VVV$&$\mathcal{L}_{5\sigma}$ (fb$^{-1}$)\\
			\hline\hline
			BP1&7.79&0.06&38.06&0.00&0.00&0.00&0.58&0.00&0.00&$6.73 \sigma$ (55)\\
			BP2&0.10&1.03&0.00&0.00&0.00&0.00&0.37&0.00&0.00&$0.92 \sigma$ (2937)\\
			BP3&0.18&0.79&3.43&4.50&0.00&0.00&0.58&0.00&0.00&$2.89 \sigma$ (299)\\
			BP4&0.00&0.45&0.00&0.01&0.00&0.00&0.74&0.01&0.00&$0.42 \sigma$ ($\gg3000$)\\
			\hline\hline
		\end{tabular}
		\caption{ The number of event combination of $\tau$-jet pair at 100 fb$^{-1}$ integrated luminosity, where $\tau$-jets are within $\pm 10 $ GeV of  $a_1$ mass peak in a final state comprised of $\geq 2\tau-\rm{jets} + \geq 2b-\rm{jets} + \geq 1 \ell$. The significance (required luminosity for $5 \sigma$ discovery) for the benchmark points are given in the last column. }\label{tautaujja}
	\end{center}
\end{table}

\begin{figure}[]
\begin{center}
\mbox{
\hspace{-1cm}
\includegraphics[width=0.7\linewidth]{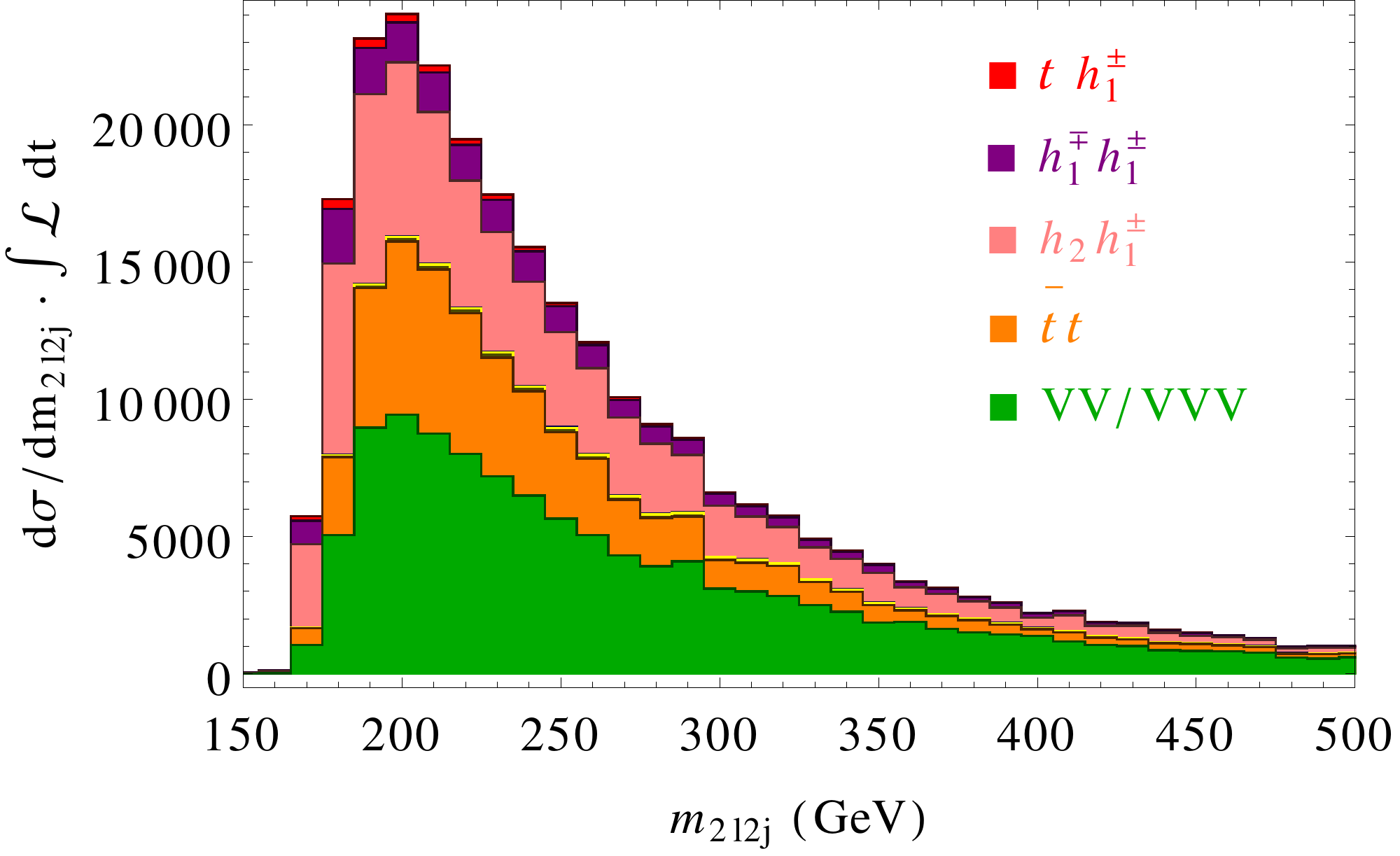}}
\caption{Invariant mass distribution $m_{2\ell2j}$ for BP4 and the dominant SM backgrounds at 100 fb$^{-1}$ of integrated luminosity. The number of events for the signals $h_1^\pm h_1^\mp$, $h_2 h_1^\pm$ and $t h_1^\pm$ are multiplied by a factor of 50 for the sake of the presentation.}\label{invmassbp4}
\end{center}
\end{figure}

Finally we focus on the charged Higgs mass reconstruction that is feasible with current run of LHC. Figure~\ref{invmassbp4} shows the invariant mass distribution of di-lepton and di-jet, \textit{i.e.} $m_{\ell\ell jj}$, where the di-leptons are coming from the $Z$ boson and are selected $\pm 5$ GeV of $Z$ mass peak of the di-lepton invariant mass distribution ($m_{\ell \ell}$)  and di-jets are coming from the $W$ boson, which are selected when they fall within $\pm10$ GeV of the di-jet invariant mass distribution $m_{jj}$. We can see that for BP4, it is possible to achieve the reconstructed charged Higgs mass peak via $h^\pm_1 \to Z W^\pm$ mode. Table~\ref{lljj} shows the reconstructed event combinations within $\pm 10$ GeV of the charged Higgs mass peak for the benchmark points and the corresponding total SM background numbers. It is clearly seen that only for BP4 a discovery of $5\sigma$ can be achieved below 1000 fb$^{-1}$ (712)  of integrated luminosity. For the rest of the points one needs very high luminosity run of LHC ($\gg 3000$ fb$^{-1}$). Thus a perfectly triplet-like singly charged Higgs boson can be easily probed via $ZW^\pm$ decay modes which is not possible for a doublet-like or a mixed charged Higgs boson at the LHC run-I. 
%%%%%%%%%%%%%%%%%%%%%%%%%
%%%%%%%%%%%%%%%%%%%%%%

\begin{table}[h]
	\begin{center}
		%\hspace*{-1.0cm}
		\renewcommand{\arraystretch}{1.3}
		\begin{tabular}{||c||c|c|c|c|c||c|c|c|c||c||}
			\hline\hline
			\multirow{2}{*}{}&\multicolumn{5}{|c||}{Signals}&\multicolumn{4}{|c||}{Backgrounds}&Significance\\
			\cline{2-10}
			&$h_1^\pm h_1^\mp$&$h_2 h_1^\pm$&$a_2 h_1^\pm$&$t h_1^\pm$&$tb h_1^\pm$&$t\bar t$&$t\bar t\, V$&$t\,Z\,W^\pm$&$VV/VVV$&$\mathcal{L}_{5\sigma}$ (fb$^{-1}$)\\
			\hline\hline
			BP1&1.09&0.00&8.78&0.00&0.00&29.44&20.69&4.36&450.65&$0.43 \sigma$ ($\gg3000$)\\
			BP2&0.77&2.03&0.00&0.00&0.00&9.81&15.52&3.49&87.24&$0.26 \sigma$ ($\gg3000$)\\
			BP3&0.00&0.57&0.14&1.50&0.00&9.81&18.00&4.56&148.70&$0.16 \sigma$ ($\gg3000$)\\
			BP4&5.83&31.12&0.00&1.27&0.00&19.62&15.26&3.39&339.84&$1.87 \sigma$ (712)\\
			\hline\hline
		\end{tabular}
		\caption{The number of event combination for $m_{\ell \ell jj}$ at 100 fb$^{-1}$ integrated luminosity, where $\ell$ are within $\pm 5 $ GeV of  $Z$ mass peak and normal jets are within $\pm 10 $ GeV of   $W^\pm$ mass peak.}\label{lljj}
	\end{center}
\end{table}

\section{Distinguishing from other extended scenario}\label{disting}

In the previous sections we have discussed the theoretical and phenomenological aspects of a charged Higgs boson in the context of the TNMSSM. This model is characterized by the presence of a light pseudoscalar in the spectrum, which allows the decay channel $h_1^\pm \to a_1 W^\pm$, as well as by the triplet-like decay channel $h_1^\pm \to Z W^\pm$. The light pseudoscalar , together with the $h^\pm \to a W^\pm$ decay mode, is also present in NMSSM \cite{ellwanger} and its phenomenology has been well studied for a doublet-like charged Higgs boson \cite{Coleppa,Kling,pbsnkh}. Similarly the presence of $ZW^\pm$ decay is signaling a triplet-like charged Higgs boson which breaks custodial symmetry \cite{quiros,TESSM}. However, TNMSSM gives an opportunity to have both the decay modes. This enables us to straightway separate the models from completely doublet-type charged Higgs and completely triplet type charged Higgs bosons. 

We considered various benchmark points in order to probe these two different decay modes of the lightest charged Higgs boson and it has been found that both $h^\pm \to a W^\pm$ and  $h_1^\pm \to Z W^\pm$ are highly improbable. Confronting Table \ref{bps} and Table \ref{Hcrosssec} we see that, apart from the pair production cross section, the relevant production channel for a triplet charged Higgs boson is either $a_2$ or $h_2$. The associated production cross section for BP1 and BP4 is $\sim 300$ fb because of the mass degeneracy between $h_1^\pm$ and $a_2$/$h_2$ respectively.

\begin{figure}[thb]
\begin{center}
\mbox{\subfigure[]{
\includegraphics[width=0.48\linewidth]{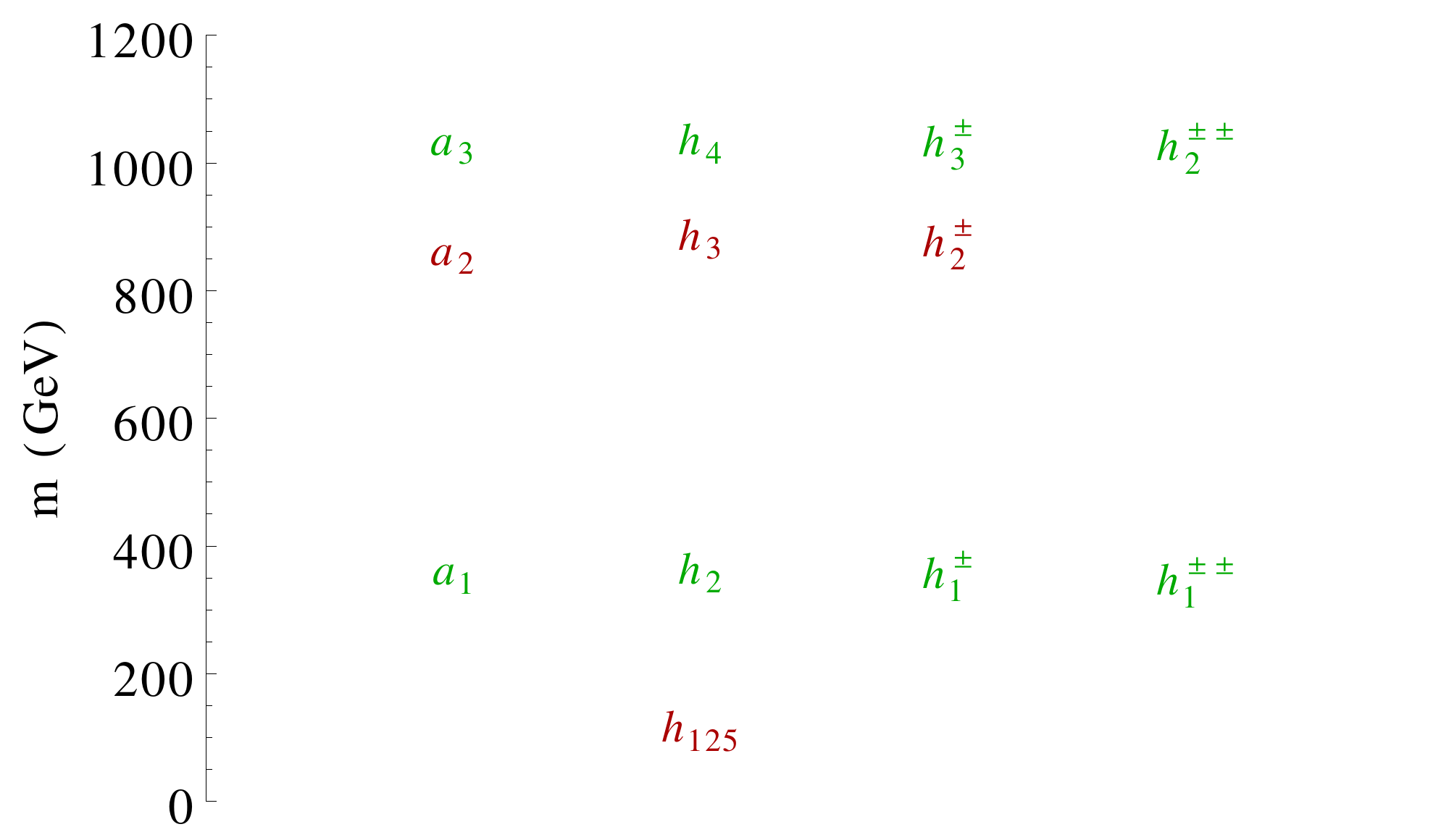}}
\subfigure[]{
\includegraphics[width=0.48\linewidth]{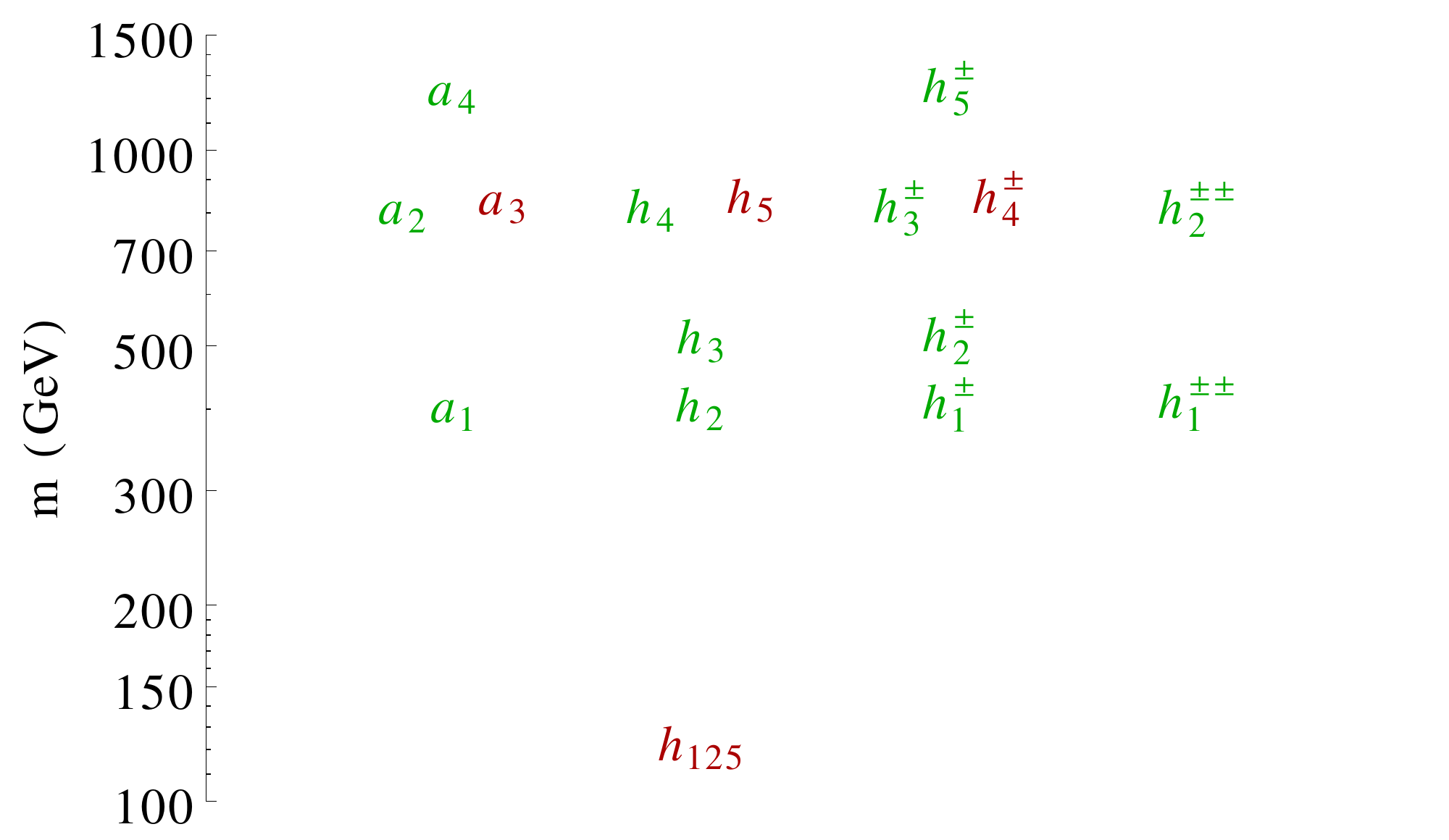}}}
\caption{A typical mass hierarchy of the scalar sector, with the doublets in red
and the triplet Higgs bosons in green color. We have considered an extension of the MSSM with a $Y=\pm1$ triplet (a) and the custodial limit of an extension of the MSSM with both $Y=0$ and $Y=\pm1$ triplets (b).}\label{cartoonT}
\end{center}
\end{figure}

The tendency of the gauge representation to group in the same mass shell scalar, pseudoscalar and charged Higgs bosons was pointed out recently \cite{OurCharged}. Here we want to emphasize that this is true even in the case of other possible triplet extensions of the MSSM. In Figure \ref{cartoonT} we present the scalar mass spectrum for two different triplet extensions of the MSSM. In Figure \ref{cartoonT} (a) we have considered the extension of the MSSM with a $Y=\pm1$ triplet superfield \cite{agashe}, whereas in Figure \ref{cartoonT} (b) the case where both the $Y=0$ and $Y=\pm1$ triplet superfields (custodial triplets) are present \cite{quiros}. The phenomenology of the $Y=\pm1$ triplet was studied in \cite{MFrank} whereas the model with triplets in the custodial symmetric limit was analyzed in \cite{Chiang}. We have selected the two sample points scanning over the parameter space and requesting the presence the doublet-type lightest scalar Higgs boson $h_1\equiv h_{125}$ with a tree-level mass $\sim125$ GeV. In both the case there are two doubly-charged charged Higgs bosons in the spectrum, because of the presence of the $Y=\pm1$ triplets. We can see that these doubly-charged states are degenerate in mass with one of the triplet neutral scalars, pseudoscalars and one of the triplet-like singly-charged Higgs bosons. Such scenarios thus claim to have  one doubly charged Higgs boson in the similar mass range of the triplet-like singly charged Higgs boson and finding both will surely shed light on the existence of the multiple $SU(2)_L$ triplets in the spectrum including the one with non-zero hypercharge. The doubly charged Higgs boson phenomenologies are independently studied in \cite{quiros,agashe,MFrank,Chiang}. Thus finding a triplet-like singly charged Higgs boson and one neutral scalar ($a_2/h_2$) with the same mass but no doubly charged Higgs bosons is a proof of existence of a $Y=0$ triplet in the spectrum. However,  existence of mass degenerate triplet-like doubly charged Higgs boson with or without additional neutral scalars along with the triplet-like singly charged Higgs boson in the mass spectrum surely tells about the existence of multiple $SU(2)_L$ triplets, \textit{i.e.} $Y=\pm 1$ and $Y=0, \pm 1$ respectively The search modes discussed in this article separates doublet- and triplet-like singly charged Higgs bosons with standard or non-standard decay modes. On top of that this mass degeneracy information along with finding or not a doubly charged Higgs boson will give us addition handle to pin down about the others gauge representation in the Higgs potential that plays a crucial role in spontaneous symmetry breaking.  
%%%%%%%%%%%%%%%%%%%%%%%%%
%%%%%%%%%%%%%%%%%%%%%%

\section{Conclusions}\label{concl}
In this article we prescribe some search modes for light singly charged Higgs boson via which we can shed light on the gauge representation of the charged Higgs bosons. A triplet-like charged Higgs boson does not couple to fermions, thus neither it is possible to produce it via fermionic modes  nor it decays in the fermionic modes, \textit{i.e.} $\tau \nu$ and/or $tb$. However a triplet-charged Higgs boson couples to $Z W^\pm$, which gives rise to additional production and decay modes. We explore these features in order to separate a triplet-like singly charged Higgs boson from a doublet one,
even in the presence of a light-pseudoscalar, which gives rise to an additional decay mode, \textit{i.e.} $a_1 W^\pm$.

For an example we have analysed signatures of a supersymmetric extension of the SM, characterized by an extra $Y=0$ Higgs triplet and a SM gauge singlet, in view of the recent and previous Higgs data. We choose different benchmark points which represent 
different decay modes preferred either by a triplet-like or by a doublet-like charged Higgs boson. BP3 represents a completely doublet-like charged Higgs boson and BP4 represents a  completely triple-like charged Higgs bosons. In the case of BP1, $a_1 W^\pm$ mode is the most dominant, whereas in BP2 we have mixed scenario. We see that a BP1 like scenario with dominant decay mode in $a_1 W^\pm$ can be probed at the LHC with 14 TeV of ECM with very early data of $\sim 43$ fb$^{-1}$ integrated luminosity via $2b+2\tau +1\ell + m_{jj}\sim m_W$ final state.

The discovery of such light pseudoscalar can be achieved with very early data of $\sim 55$ fb$^{-1}$ and this would be certainly a signal in favour of an extended Higgs sectors. The NMSSM does have such light pseudoscalar but does not have any extra charged Higgs bosons compared to the MSSM, while the TNMSSM has an extra triplet-like charged Higgs boson. This possibility changes the direct bounds derived from searches for a charged Higgs at the LHC, as well as the indirect bounds on flavour. These changes are due to the doublet-triplet mixing in the charged Higgs and chargino sectors of the triplet extended model \cite{tripch}.

Triplet-like charged Higgs boson and its decay mode to $ZW^\pm$ can be probed 
via $3\ell + 1\tau $ with an early data of $\sim 54$ fb$^{-1}$ of integrated luminosity.
The charged Higgs mass can be reconstructed with relatively larger data of $\sim 712$ fb$^{-1}$. Finding such triplet-charged state would clearly be proof of the existence of 
higher representation of $SU(2)_L$ in the Higgs potential. We also present the sensitivity in other possible search modes. 
 
 The existence of non-zero hypercharge triplet would require to have one doubly charged Higgs boson degenerate with the singlet triplet-like charged Higgs boson. Search for those doublet-like states can one tell us about the different possible triplet representation with non-zero hypercharge.

%%%%%%%%%%%%%%%%%%%%%%%%%%%%%%%%%

\end{document}